\documentclass[11pt,a4paper]{article}
\usepackage[makeroom]{cancel}
\usepackage{latexsym}
\usepackage{longtable}
\usepackage{epsfig}
\usepackage{graphicx,bbm,psfrag}
\usepackage{amssymb,amsmath,amsthm,booktabs,mathtools}
\usepackage{bm}
\usepackage{color}
\usepackage{dutchcal}

\usepackage[backend=biber,style=phys,sorting=none,date=year,isbn=false,url=false,doi=false,eprint=false]{biblatex}
\AtEveryBibitem{
 \clearfield{eprint}
 \clearfield{isbn}
 \clearfield{issn}
 \clearlist{location}
 \clearfield{month}
 \clearfield{number}
 \clearfield{issue}
 \clearfield{language}
}

\definecolor{purple_nice}{rgb}{0.4,0.2,0.7}
\definecolor{fuel_blue}{RGB}{42,162,185}
\definecolor{YInMn_blue}{RGB}{46, 80, 144}
\definecolor{ultramarine}{RGB}{63, 0, 255}
\definecolor{KLEIN_blue}{rgb}{0, 0.18, 0.65}
\usepackage[linktocpage=true]{hyperref}
\hypersetup{
    colorlinks=true,
    linkcolor=YInMn_blue,
    citecolor=ultramarine,
    filecolor=fuel_blue,
    urlcolor=KLEIN_blue,
}

\newtheorem{theorem}{Theorem}[section]
\newtheorem{rema}{Remark}[section]

\newtheorem{defi}[rema]{Definition}
\newtheorem{lemma}[theorem]{Lemma}
\newtheorem{corol}[theorem]{Corollary}

\newcommand{\bc}{\begin{center}}
\newcommand{\ec}{\end{center}}
\def\ba#1{\begin{array}{#1}\displaystyle}
\newcommand{\ea}{\end{array}}

\newcommand{\beq}{\begin{equation}}
\newcommand{\eeq}{\end{equation}}
\newcommand{\beqa}{\begin{eqnarray}}
\newcommand{\eeqa}{\end{eqnarray}}
\newcommand{\no}{\nonumber}
\newcommand{\n}{\nonumber\\}
\newcommand{\bi}{\begin{itemize}}
\newcommand{\ei}{\end{itemize}}

\def\lt#1{\left#1}
\def\rt#1{\right#1}

\def\b#1{\bar{#1}}
\def\frc#1#2{\frac{#1}{#2}}

\def\bs#1{\boldsymbol{#1}}

\newcommand{\I}{\mathbb I}

\newcommand{\p}{\partial}

\newcommand{\N}{{\mathbb{N}}}
\newcommand{\R}{{\mathbb{R}}}

\newcommand{\ep}{\epsilon}
\newcommand{\varep}{\varepsilon}

\newcommand{\ri}{{\rm i}}

\newcommand{\dd}{{\rm d}}
\newcommand{\1}{{\bf 1}}
\DeclareMathOperator{\sgn}{sgn}

\newcommand{\halmos}{\rule{1ex}{1.4ex}}
\newcommand{\eproof}{\hspace*{\fill}\mbox{$\halmos$}}

\newcommand{\ttbar}{$T\bar{T}$}

\begin{document}

\begin{center}
{\Large {\sc Generalised $T\bar T$-deformations\\[0.1cm] of classical free particles}}

\vspace{1cm}

{\large Benjamin Doyon$^1$, Friedrich H\"ubner$^1$, Takato Yoshimura$^{2,3}$}

\vspace{0.2cm}
$^1$Department of Mathematics, King’s College London, Strand, London WC2R 2LS, U.K.\\
$^2$All Souls College, Oxford OX1 4AL, U.K.\\
$^3$Rudolf Peierls Centre for Theoretical Physics, University of Oxford,
1 Keble Road, Oxford OX1 3NP, U.K.
\ec

Deformations of many-body Hamiltonians by certain products of conserved currents, referred to as $T\b T$-deformations, are known to preserve integrability. Generalised $T\b T$-deformations, based on the complete space of pseudolocal currents, were suggested [B. Doyon, J, Durnin, T. Yoshimura, Scipost Physics 13, 072 (2022)] to give rise to integrable systems with arbitrary two-body scattering shifts, going beyond those from known models or standard CDD factors.
However, locality properties were not clear. We construct explicit generalised $T\b T$-deformations of the system of classical free particles. We show rigorously that they are Liouville integrable Hamiltonian systems with finite-range interactions. We show elastic, factorised scattering, with a two-particle scattering shift that can be any continuously differentiable non-negative even function of momentum differences, fixed by the \ttbar-deformation function. We show that the scattering map (or wave operator) has a finite-range property allowing us to trace carriers of asymptotic momenta even at finite times — an important characteristics of many-body integrability. We evaluate the free energy and prove the thermodynamic Bethe ansatz with Maxwell-Boltzmann statistics, including with space-varying potentials and in finite and infinite volumes. We give equations for the particles' trajectories where time appears explicitly, generalising the contraction map of hard rod systems: the effect of generalised $T\b T$-deformations is to modify the local metric perceived by each particle, adding extra space in a way that depends on their neighbours. The systems generalise the gas of interacting Bethe ansatz wave packets recently introduced in the Lieb-Liniger model. They form a new class of models that, we believe, most clearly make manifest the structures of many-body integrability.


\tableofcontents

\section{Introduction}

Recently, a very special type of deformations of many-body systems has been introduced in the context of 1+1-dimensional quantum field theory (QFT) \cite{Cavagli2016,smirnov_space_2017}. These so-called ``$T\bar T$-deformations" are deformations of the Hamiltonian by bilinear terms involving local conserved currents. They fundamentally affect the dynamics, but, if the system is completely integrable, they preserve its integrability. They are so referred to, because one of the simplest example is a deformation, in conformal field theory, by a product of the holomorphic, $T$, and anti-holomorphic, $\b T$, components of the stress tensor \cite{Zamolodchikov:2004ce}. \ttbar-deformations have been applied to systems of different kinds, including quantum chains \cite{Pozsgay2020,PhysRevE.104.044106,PhysRevE.104.064124,PhysRevLett.131.037101} and other quantum and classical non-relativistic models \cite{cardy1809tt,esper2021tt,cardy2022t,jiang2022mathrm}. Their effects on the thermodynamics have been studied \cite{Cavagli2016,hernandez-chifflet_flow_2020,10.21468/SciPostPhys.13.3.072}, and very recently form factors, which describe correlation functions, have been proposed \cite{castroalvaredo2023form,castroalvaredo2023form2,hotrepresentation2}. See e.g.~the review \cite{jiang2021pedagogical}. 

In order to understand the importance of \ttbar-deformations, let us recall how the structure of scattering plays a crucial role in integrable many-body systems. Take for instance a classical Hamiltonian system of particles in one dimension \cite{hubacherClassical1989}, with interaction between particles that is translation-invariant and short-range, and without bound states. Naturally, it is a general property of such systems, integrable or not, that at large times, particles typically separate out and follow straight trajectories determined by ``asymptotic coordinates": constant asymptotic momenta (their slopes) and impact parameters (their spatial shifts). The scattering map, often called ``wave operator" -- the symplectic transformation from real to asymptotic coordinates, either ``in" or ``out" -- is a canonical transformation that trivialises the dynamics. Thus, the asymptotic momenta form a set of conserved quantities in involution, much like in Liouville-Arnold integrability  \cite{arnoldbook}. However, the existence of such a trivialising transformation certainly does not, in fact, imply integrability!

Two effects occur from integrability. First, in integrable models many-body scattering is very constrained: it is elastic and factorised. That is, the sets of asymptotic in- and out-momenta is the same (elastic scattering), and the full scattering shift incurred by each particle is the sum of successive two-body scattering shifts, as if many-particle scattering were a succession of well separated two-particle events (factorised scattering). Elastic scattering follows from factorised scattering by energy-momentum conservation at two-particle events; and under certain conditions on the interaction potential, factorised scattering can also be shown to follow from elastic scattering \cite{hubacherClassical1989}. Second, the scattering map is expected to have special properties. In addition to asymptotic momenta being differentiable functions of phase-space coordinates, they satisfy a short-range condition, guaranteeing that it is somehow possible to ``trace" the carrier of a given asymptotic momentum throughout time, from its in- to its out-trajectory. This is closely related to the crucial condition, for integrable systems with unbounded trajectories, that Liouville conserved quantities be {\em extensive}, essentially they must possess densities acting on short spatial ranges. Such ``short-range Liouville integrability" implies elastic and factorised scattering by Parke's  argument, originally made in QFT \cite{parke_absence_1980} but more generally applicable \cite{doyon_generalised_2019,doyon_lecture_2019}. The short-range property of asymptotic momenta is however in general difficult to establish, besides simple models such as hard rods \cite{Boldrighini1983}. Similar concepts hold in integrable PDEs, QFTs, quantum chains, cellular automata (e.g. Rule 54 \cite{Buca_2021} and box-ball \cite{2020cs} systems offer good examples of the scattering theory), etc.

This scattering structure is important. For instance, it is well known that the thermodynamics of completely integrable many-body systems is determined solely by the two-body scattering shift function $\varphi(\theta-\theta')$ (for incoming momenta $\theta,\theta'$)\footnote{The space of asymptotic particles is taken for simplicity to be characterised by asymptotic momenta $\theta$ only, and we assume a dependence on difference of momenta. But the TBA and GHD are applicable more generally, including to models with more internal degrees of freedom and higher-dimensional spectral space.}, and takes the universal form of the thermodynamic Bethe ansatz (TBA)  \cite{yangyang,Zamolodchikov1990,takahashi2005thermodynamics,mossel2012generalized,doyon_lecture_2019}. Similarly, the emergent large-scale dynamics takes the universal form of generalised hydrodynamics (GHD) \cite{PhysRevX.6.041065,PhysRevLett.117.207201,doyon_lecture_2019,spohn2023hydrodynamic}.
Yet, only very specific two-body scattering shifts (classical) or amplitudes (quantum), as functions of incoming momenta, are known to correspond to explicit integrable models with short-range interactions. For instance, in classical particle systems, hard rods have constant two-body shift function $\varphi(\theta-\theta')=-a$ given by the length of the rod $a$ (generalised to rods of different lengths \cite{ferrari2022hard,cardy2022t}), and for the Toda model $\varphi(\theta-\theta') = 2\log|\theta-\theta'|$ (see \cite{Moser1975} for an analysis of the scattering theory). 
What is the full space of two-body scattering shifts consistent with integrable short-range interactions, and can we prove the TBA and more subtle scattering properties?

In QFT, $T\b T$-deformations were shown to give rise to arbitrary products of CDD factors in the scattering amplitude \cite{smirnov_space_2017}. Further, it was recently realised \cite{10.21468/SciPostPhys.13.3.072} that one could adapt the insight coming from non-equilibrium quantum dynamics \cite{Ilievski_2016,Doyon2017,de2022correlation} and consider the complete space of extensive (or pseudolocal) conserved quantities. What we refer to here as ``generalised \ttbar-deformations" are the deformations involving the associated densities and currents. They give rise to {\em arbitrary} deformations of scattering amplitudes, going beyond standard CDD factors \cite{10.21468/SciPostPhys.13.3.072}, thus covering a large family of integrable models; this is technically because the extensive conserved quantities are spanned by the charges measuring the density of asymptotic momenta \cite{PhysRevB.95.115128,de2022correlation}. This gives the hope of a better understanding of the space of integrable many-body systems. Locality properties of generalised \ttbar-deformations are however unclear. Do generalised $T\b T$-deformations of short-range many-body systems still have short-range interactions? Can we construct them explicitly?

In this paper, we address these questions in a rigorous fashion. We construct the explicit generalised $T\b T$-deformations of the system of classical free particles. We consider, in fact, a more general form than that used in the literature, based on a deformation function $w(x,\theta)$ of both space and momentum (within a large family of functions). We exhibit the generating function that maps real to asymptotic coordinates for the \ttbar-deformed systems; it is closely related to the form of the Bethe ansatz wave function. We  show that the deformations are short-range Liouville integrable Hamiltonian systems, with interactions whose spatial range is in fact finite (in a specific way that we will explain). The interaction potential is differentiable and particle-symmetric, but depends in general on momenta -- this allows us to elude the theorem according to which integrable interaction potentials cannot be finite-range \cite{khimchenko1984}.

We show that the \ttbar-deformed systems satisfy elastic, factorised scattering, with a two-particle scattering shifts $\varphi(\theta-\theta')$ that can be chosen as any continuously differentiable, real, non-negative, symmetric functions of the difference of momenta; hence this gives credence to the claims made in \cite{10.21468/SciPostPhys.13.3.072}. The microscopic Hamiltonian depends on the full function $w(x,\theta)$, while the scattering shift only depends on its integral over $x\in\R$. In any scattering process, particles fully exchange their positions (the ``go-through" picture of \cite{cardy2022t}): each is a ``tracer" of its asymptotic momentum. The systems are shown to have a free energy given by the TBA with Maxwell-Boltzmann statistics, including with space-varying generalised  potentials --  the local density approximation holds -- and even at finite volumes.

We also show that the scattering map has the properties of many-body integrability. Not only the map to asymptotic momenta is a continuously differentiable function, but also we show that it satisfies a short-range property, allowing us to estimate the positions of carriers of asymptotic momenta even at finite times. That is, if a set of particles is separated enough from the rest, their positions and momenta fully determine their asymptotic momenta, and these asymptotic momenta are simply those one would obtained in a ``time-of-flight" experiment, where the particles are taken out of the system and let to expand in the vacuum. The time-of-flight argument is one way of justifying the emergence of the GHD equations based on the scattering picture of the thermodynamics (see e.g.~\cite{doyon_lecture_2019,doi:10.1126/science.abf0147,doyon2023ab}).


The trajectories of the generalised $T\b T$-deformed systems can be written as equations where time appears explicitly. This is similar to the contraction map of the system of hard rods \cite{Spohn1991} (and its generalisation to rods of different lengths \cite{ferrari2022hard,cardy2022t}). The system of hard rods is known to be obtained from certain $T\b T$-deformations, which change the lengths of fundamental particles \cite{cardy2022t}. Here, instead, it is a change of the metric perceived by each particle: more space (because $\varphi(\theta-\theta')\geq 0$) is available to a particle whenever other particles are near, in which it then travels freely. Therefore, generalised $T\b T$-deformations change the local space available to particles. The local metric where particles travel freely becomes, at large scales, the metric shown to map the GHD equations to the free-particle Liouville equation \cite{Doyon2018}, and more generally this makes rather explicit the geometric interpretation of \ttbar-deformations \cite{Conti2019}. We address the emergence of the GHD equations from our system of particles in \cite{DHY23b}. The trajectories generalise the gas of interacting semi-classical Bethe ansatz wave packets recently introduced in the Lieb-Liniger model \cite{doyon2023ab}.

As far as we are aware, this is a new family of classical integrable systems, never considered before; however there may be connections with sine-Gordon soliton trajectories \cite{babelon1993sine,babelon1996quantization,babelon1997form}.

The paper is organised as follows. In Section \ref{sectmain} we express our main theorem, Theorems \ref{thsummary}, \ref{thsummaryttbar} and \ref{thsummaryTBA}, and provide a discussion of their implications and meaning. In Section \ref{sectProperties}, we provide accurate definitions of properties of many-body classical particle systems (that we will show hold for the systems we introduce): Liouville integrability, finite-range interactions, Elastic and factorised scattering, tracer dynamics, generalised $T\b T$-deformations, and the thermodynamic Bethe ansatz. Finally, in Section \ref{sectproofs} we provide the proofs and in Section \ref{sectconclu} a short conclusion.

\section{Main results and discussion}\label{sectmain}

In this section we present our setup and main results. The results are about certain trajectories in canonical phase space $\R^{2N}$ ($N\in\N=\{0,1,2,\ldots\}$), which are defined by a certain transformation of coordinates. We state that they correspond to a Liouville integrable Hamiltonian evolution with finite-range interaction, which has the property of elastic, factorised scattering, which is a generalised $T\b T$-deformation of the system of free particles, and whose thermodynamics is given by the classical thermodynamic Bethe ansatz (TBA), with Maxwell-Boltzmann statistics. All these notions, that we show hold for this Hamiltonian system, are notions that are well understood at least informally, and have been studied in various setups (such as classical mechanics, QFT, Bethe ansatz integrability, PDEs, soliton gases, etc.). In order express our results unambiguously, we provide precise definitions of these notions in the setup of many-particle classical systems in Section \ref{sectProperties}. 
%
%

Let $\psi\in C^2(\R^2)$ and $\ep>0$. The following properties are assumed: for all $(x, \theta)\in\R^2$,
\beq\label{condpsioverview}
	\psi_{x\theta}(x,\theta)\geq 0,\quad \psi(-x,-\theta) = \psi(x,\theta),\quad 
	\psi_x(x,\theta) = 0\ \mbox{whenever $|x|\geq\ep$}
\eeq
where $\psi_x = \p\psi/\p x$, $\psi_\theta = \p\psi/\p \theta$, etc. Let us define $\varphi(\theta)$ as
\beq\label{defphioverview}
	\varphi(\theta)
    = \int_{\R} \dd x\,\psi_{x\theta}(x,\theta) =
    \psi_\theta(X,\theta)
	- \psi_\theta(-X,\theta)\quad \mbox{(every $X>\ep$).}
\eeq
Note that $\varphi(\theta)\geq 0$ for all $\theta\in\R$, that $\varphi\in C^1(\R)$, and that $\varphi(\theta) = \varphi(-\theta)$.

Below we use the boldface notation for vectors $\bs y\in\R^N$, with components $y_i$ for $i=1,2,\ldots,N$, etc., and we denote $\R^{2N}_{\neq}:=\{(\bs y,\bs \theta)\in\R^{2N}:\theta_i\neq \theta_j\;\forall\; i\neq j\}$.

The main object is the map $\omega : \R^{2N} \ni (\bs y,\bs \theta) \mapsto (\bs x,\bs p)\in\R^{2N}$ defined as the solution to the following system of nonlinear equations:
\begin{align}\label{cba1}
    y_i &=
    x_i + \sum_{j\neq i}\psi_\theta(x_i-x_j,\theta_i-\theta_j)\\ \label{cba2}
    p_i &= \theta_i + \sum_{j\neq i}
    \psi_x(x_i-x_j,\theta_i-\theta_j).
\end{align}
Specifically, one solves \eqref{cba1} for $\bs x$, and then \eqref{cba2} gives $\bs p$. The variables $(\bs x,\bs p)$ are to be interpreted as the canonical coordinates of $N$ particles, each lying in one-particle phase space $\R^2$, with canonical Poisson structure $\{x_i,p_j\}= \delta_{ij}$, etc. We also define the evolution one-parameter group $\tau_t:(\bs y,\bs \theta) \mapsto (\bs y+\bs \theta t,\bs\theta)$ for every $t\in \R$, and the time evolution on the canonical coordinates is obtained by the composition $\omega\circ\tau_t$. In this way \eqref{cba1}, with $\bs y(t) = \bs y + \bs\theta t$ and $\bs\theta$ constant, defines the trajectories $t\mapsto (\bs x(t),\bs p(t)) = \omega(\tau_t(\bs y,\bs \theta))$ of our model of classical interacting particles. We will refer to these as {\em semiclassical Bethe systems} (for reason that we explain below).

\subsection{Finite-range integrable models with arbitrary scattering shift}\label{ssectarbitrary}

We will show that this dynamics has interesting properties: the trajectories indeed exist and are unique by the above definition, they are generated by a Hamiltonian that is Liouville integrable and that has finite-range interaction, and the many-body scattering of the particles is elastic, and factorises into two-body scattering events.
\begin{theorem}[Dynamics] \label{thsummary} {\ }
\begin{enumerate}
\item[a.] {\em (Trajectories).} For every $\bs\theta\in\R^N$, the map $\bs y\stackrel{\mbox{\eqref{cba1}}}\mapsto \bs x$ is a $C^1$ diffeomorphisms of $\R^N$, and the function $\bs x\stackrel{\mbox{\eqref{cba2}}}\mapsto \bs p$ is continuously differentiable. Thus the trajectories $t\mapsto (\bs x(t),\bs p(t))$ exist, are unique, and are continuously differentiable functions of $t$. For every $(\bs y,\bs\theta)\in\R^{2N}_{\neq}$, there is a time $T>0$ such that for all $|t|\geq T$, trajectories are linear with $\dot{\bs x}(t) = \bs p(t) = \bs \theta$. Thus $\bs\theta$ are asymptotic momenta. Finally, the map $\omega$ is a $C^1$ diffeomorphism. (See the full statement in Theorem \ref{thetrajectoriesfull}.)
\item[b.] {\em (Liouville integrability and finite-range interaction).} The Hamiltonian function $H_\psi$ defined by
\beq\label{Hpsi}
    H_\psi(\bs x,\bs p) = \sum_{i=1}^N \frc{\theta_i(\bs x,\bs p)^2}2 =:
    \sum_{i=1}^N \frc{p_i^2}2 + V_\psi(\bs x,\bs p),\qquad
    H_\psi\in C^1(\R^{2N})
\eeq
where $V_\psi(\bs x,\bs p) = \sum_i (\theta_i(\bs x,\bs p)^2-p_i^2)/2$, generates the dynamics,
\beq
    \dot x_i(t) = \{x_i(t),H_\psi\},\quad
    \dot p_i(t) = \{p_i(t),H_\psi\}.
\eeq
It is Liouville integrable with finite-range interaction (Definition \ref{defilocal}). In particular, if a set $J$ of positions $\{x_j: j\in J\}$ is such that $|x_i-x_j|>\ep$ for all $i\in \{1,\ldots,N\}\setminus J$, then $\theta_j(\bs x,\bs p) = \theta_j(\bs x_J,\bs p_J)$ for all $j\in J$, where the function $\theta_j$ restricted to $\bs x_J=(x_j)_{j\in J}$, $\bs p_J=(p_j)_{j\in J}$ is obtained from writing \eqref{cba2} on that subset only. Thus asymptotic momenta $\theta_j$'s for any cluster $J$ of particles separated enough from other particles, are independent from the coordinates of other particles, and are obtained by a time-of-flight thought experiment for the cluster $J$ put in the vacuum. See Fig \ref{fig:tof}.
\item[c.] {\em (Elastic, factorised scattering).} $H_\psi$ satisfies the property of elastic and factorised scattering, and is a finite-range tracer dynamics, with two-body scattering shift $\varphi(\theta-\theta')$ for incoming momenta $\theta,\,\theta'$ (Definitions \ref{defiscattering} and \ref{defifinitetracer}).
\end{enumerate}
\end{theorem}
Thus the semiclassical Bethe systems are ``nice" systems of interacting particles, albeit with quasi-potentials $V_\psi(\bs x,\bs p)$ (``quasi" because of the dependence on momenta) whose evaluation requires us to solve a system of nonlinear equations. Note that the notion of ``tracer dynamics" (Part c of the theorem) means that in any scattering process, particles fully exchange their positions -- this is the ``go-through" picture of \cite{cardy2022t}, where each particle is a tracer of its asymptotic momentum.

We mention four points of interest.

{\bf 1.} We have obtained a large family of finite-range Liouville integrable Hamiltonian systems with elastic, factorised scattering, which admits {\em an arbitrary two-body scattering shift} as a function of difference of momenta: for any real, non-negative, symmetric and continuously differentiable function, there corresponds at least one model in the family. That is, given $\varphi\in C^1(\R)$ with $\varphi(\theta)\geq 0$ and $\varphi(\theta) = \varphi(-\theta)$ for all $\theta\in\R$, we can find $\psi(x,\theta)$ in $C^2(\R^2)$ satisfying \eqref{condpsioverview} and \eqref{defphioverview}:
\beq\label{psigenericphi}
	\psi(x,\theta) = \phi(\theta) \times
	\lt\{\ba{ll}
	\frc{I(x)}{2I(\ep)} & (|x|<\ep)\\[0.3cm]\displaystyle
	\frc12\sgn(x) & (|x|\geq \ep)
	\ea\rt.
\eeq
where
\beq\label{phase}
	\phi(\theta) = \int_0^\theta \dd\theta'\varphi(\theta'),\quad
    I(x) = \int_0^x\dd x'
    e^{-1/(\ep-x') - 1/(\ep+x')}.
\eeq
Note that this $\psi(x,\theta)$ is in fact smooth in $x$. This goes much beyond the special potentials and two-body scattering shifts that are known from specific models up to now. We also see that there is a lot of freedom in the construction: a given two-body shift {\em does not} uniquely fix the local interaction. See \cite{DHY23b} for a discussion of the relation to models of hard rods \cite{Spohn1991,cardy2022t,ferrari2022hard}.

\begin{figure}
    \centering
    \includegraphics[width=12cm]{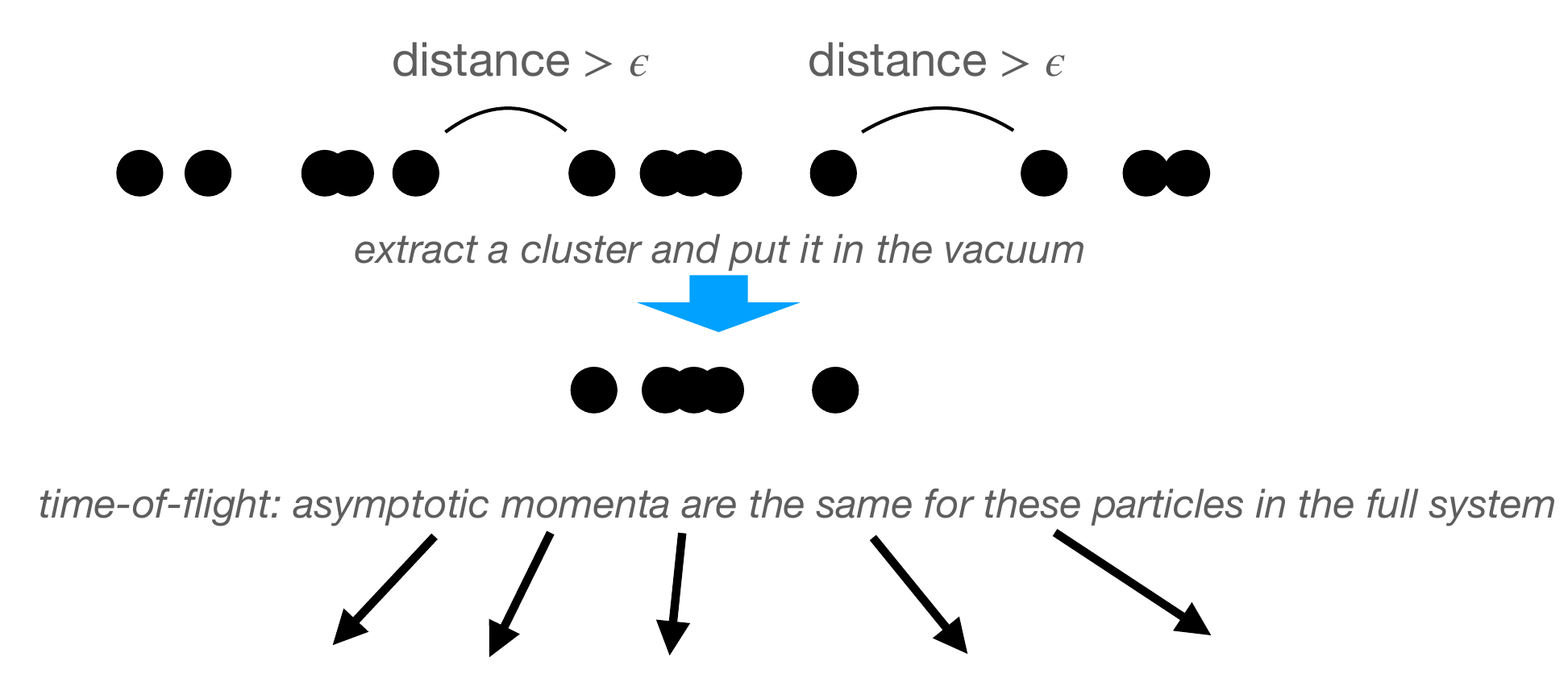}
    \caption{Finite-range interaction. Whenever a cluster of particles is a distance greater than $\epsilon$ from other particles, it does not interact with other particles. Extracting the cluster, putting it in the vacuum, and letting it expand by the dynamics restricted to these particles, the asymptotic momenta emerging at large times are exactly those that would have emerged at large times for these particles in the full system. Thus, the asymptotic momenta can be traced -- they are evaluated from the positions and momenta of a cluster and spatially lie there -- even at finite times and non-zero densities. The set of asymptotic momenta spatially lying in a large region, at the Euler scale, characterises the ``fluid cell" of this region -- this is at the basis of GHD.}
    \label{fig:tof}
\end{figure}

{\bf 2.} Part a says that the asymptotic momenta are $C^1$ functions of the particles' phase-space coordinates. In fact, the full out/in-scattering map $S^\pm:(\bs x,\bs p)\mapsto (\bs x^\pm,\bs\theta)$,
\beq
    \bs x(0) = \bs x,\quad \bs p(0) = \bs p;\quad x_i(t) = x_i^{\sgn(t)} + \theta_i t,\quad
    p_i(t) = \theta_i \qquad (|t|>T),
\eeq
is given by $\bs\theta(\bs x,\bs p)$ defined above, and
\beq
    x_i^\pm = y_i
    - \sum_{j\neq i} 
 \psi_\theta(\pm\sgn(\theta_i-\theta_j)\times\infty,\theta_i-\theta_j)
\eeq
(see \eqref{xipiasymp}). Thus the impact parameters at both positive and negative infinite times are linear in $\bs y$ for given  $\bs\theta$. One can consider $\bs y$ as the ``better" choice of impact parameters, for which the  scattering map $\omega^{-1}:(\bs x,\bs p)\mapsto (\bs y,\bs\theta)$ is a $C^1$ diffeomorphism. Further, Part b shows that {\em the asymptotic momenta satisfy a finite-range property}: the coordinates of a cluster of particles separated from other particles fix its set of asymptotic momenta by a time-of-flight thought experiment. As discussed in the introduction, this is an important property of many-body integrable systems, which gives support to the general argument for the GHD equation based on the scattering picture (see e.g.~\cite{doyon_lecture_2019}). See Fig.~\ref{fig:tof}.

{\bf 3.} It turns out that \eqref{cba1} and \eqref{cba2} take the form of a canonical transformation
\begin{align}\label{yeq}
	y_i &= \p_{\theta_i}\Phi(\bs x,\bs \theta) \\
	p_i &= \p_{x_i}\Phi(\bs x,\bs \theta)
	\label{peq}
\end{align}
generated by
\beq\label{generatingsummary}
	\Phi(\bs x,\bs\theta) =
	\sum_i x_i\theta_i
	+\frc12\sum_{i\neq j} \psi(x_i-x_j,\theta_i-\theta_j).
\eeq
In \cite{doyon2023ab}, an ab initio derivation of the GHD equation is proposed in the Lieb-Liniger (LL) model. The derivation is based on constructing slowly-varying modulations of Bethe ansatz wave functions, and identifying them as representing gases of wave packets. A semiclassical analysis then gives the dynamics for these wave packets. It turns out that this dynamics is nothing else but \eqref{cba1}, with
\beq
	\psi(x,\theta) = \frc12 \sgn(x)\phi(\theta)
\eeq
where $\phi(\theta)$ is the scattering phase of the Bethe ansatz for the LL model. This is  \eqref{psigenericphi} in the limit $\ep\to0$. In fact, Eqs.~\eqref{yeq}, \eqref{peq} are more generally coming from a semiclassical analysis of Bethe wave packets, for the phase $\Phi(\bs x,\bs\theta)$ of the Bethe ansatz wave function, $\Psi(\bs x) = \sum_{\sigma\in S_N} e^{\ri \Phi(\bs x_\sigma,\bs\theta)}$ (where $S_N$ is the permutation group). The semiclassical arguments say that $(\bs y,\bs\theta)$ are then freely-evolving coordinates. In our models, $\sum_{i\neq j} \psi(x_i-x_j,\theta_i-\theta_j)$ replaces what occurs in the Bethe ansatz, $\frc12\sum_{i\neq j} \sgn(x_i-x_j)\phi(\theta_i-\theta_j)$.  Thus, what we refer to as the semiclassical Bethe systems \eqref{cba1}, \eqref{cba2} are indeed closely related to the Bethe ansatz of quantum integrable models, in a certain ``semiclassical" regime of slowly varying amplitudes. It is natural to ask if \eqref{generatingsummary} can be used as a Bethe ansatz phase for a finite-range integrable quantum model, a ``quantisation" of a semiclassical Bethe system; we hope to address this in future works.

{\bf 4.} Finally, there is a geometric interpretation of \eqref{cba1}: taking the differential of $y_i$ keeping coordinates $x_j,\,j\neq i$ fixed, at $t=0$, we obtain
\beq
    \dd y_i = \Big(1 +
    \sum_{j\neq i}\psi_{x\theta}(x_i-x_j,\theta_i-\theta_j)\Big)\dd x_i.
\eeq
The factor in the parenthesis on the right-hand side is strictly positive, uniformly for all $\bs x\in\R^N$. Therefore, this is a change of metric on $\R$ for particle $i$. The metric depends on the positions $x_j$ of other particles that are a distance $\ep$ or less to particle $i$ (and on their asymptotic momenta, which are fixed throughout time evolution). As the coordinate $y_i$ is freely evolving, particle $i$ can be seen as being free, but in a metric that is affected by the positions of other particles nearby. The situation here is much like that of rods with negative lengths, as the spaces where the particles evolve freely, with coordinates $\bs y$, are {\em larger} than that with the original coordinates $\bs x$. This is the explicit microscopic equivalent of the geometric ideas underlying the GHD equations \cite{Doyon2018} and \ttbar-deformations \cite{Conti2019} (see below for the connection to \ttbar-deformations)\footnote{It is also much like in Einstein's general theory of relativity, where interaction is implemented by a configuration-dependent metric. Here, though, for any given configuration, the metric felt by different particles is generically not the same (because of the dependence of the metric on the rapidities), by contrast to Einstein's general theory of relativity.}.

\begin{rema}
The restriction to scattering shifts that are functions of the difference of momenta is for simplicity, and can easily be lifted, and we believe the requirements of symmetry and of continuous differentiability can also be lifted without too much difficulty. Taking away the non-negativity requirement may involve more analysis, in particular it may be necessary to restrict the phase space in order to avoid singularities; see \cite{DHY23b} for a discussion. For $\psi(x,\theta)$, one can extend, also without too much difficulty, to functions whose derivative $\psi_x(x,\theta)$ vanish sufficiently fast as $|x|\to\infty$, instead of the last equation of \eqref{condpsioverview}.
\end{rema}

\subsection{Generalised $T\b T$ deformations of free particles}\label{ssectmainTTbar}

The equations \eqref{cba1}, \eqref{cba2} may seem rather {\em ad hoc}. In order to understand their origin, and the meaning of the Hamiltonian \eqref{Hpsi}, one route is via $T\b T$-deformations, discussed in Subsection \ref{ssectTTbar}. A generalised $T\b T$-deformation of a Hamiltonian $H_0$ is a flow of Hamiltonians $\lambda \mapsto H_\lambda$, which takes the form
\beq
	\frc{\p H_\lambda(\bs x,\bs p)}{\p \lambda} 
	= -\sum_{ij}
	\big(\dot x^\lambda_{i}(\bs x,\bs p) - \dot x^\lambda_{j}(\bs x,\bs p)\big)\,w\big(x_{i}-x_{j},\theta^\lambda_{i}(\bs x,\bs p)-\theta^\lambda_{j}(\bs x,\bs p)\big)
 \label{Hhatsummary}
\eeq
for all (say) $\lambda\in[0,1]$ and a.e.~$(\bs x,\bs p)\in\R^{2N}$, and for some continuously differentiable ``deformation function" $w(x,\theta)$. On the right-hand side, the upper index $^\lambda$ indicates that the quantities are evaluated with respect to the Hamiltonian $H_\lambda$, in particular $\dot x_i^\lambda(\bs x,\bs p) = \{x_i,H_\lambda(\bs x,\bs p)\}$ and $\theta^\lambda_{i}(\bs x,\bs p)$ are asymptotic momenta with respect to the dynamics induced by $H_\lambda(\bs x,\bs p)$. Then we say that $H_\psi$ is a generalised $T\b T$ deformation of $H_0$ if there is such a flow such that $H_\psi = H_1$. 

The above is a generalisation of the deformation by products of local conserved densities and currents that is the hallmark of $T\b T$-deformations \cite{smirnov_space_2017}. Here, as originally proposed in \cite{10.21468/SciPostPhys.13.3.072}, it is generalised to densities and currents that are parametrised by a continuous parameter: an asymptotic momentum. This generalisation is possible, as, being tracer dynamics, our family of models admit well-defined extensive conserved quantities $Q_\theta$, $\theta\in\R$ that measure the density of asymptotic particles of asymptotic momenta $\theta$, and that {\em possess local densities and associated currents}, phase-space functions $q_{\theta}(\bs x,\bs p;x)$ and $j_\theta(\bs x,\bs p;x)$ satisfying $\dot q_\theta(x) + \p_x j_\theta(x)=0$ (keeping the phase-space dependence implicit). In terms of the latter, a generalised $T\b T$-deformation takes the form
\beq\label{genTTbarsummary}
	\frc{\p H_\lambda}{\p \lambda} = \int \dd\theta\dd\alpha
	\dd x\dd x' \, w(x-x',\theta-\alpha)\big( q^\lambda_{\theta}(x)j^\lambda_{\alpha}(x')
	-j^\lambda_{\theta}(x)q^\lambda_{\alpha}(x')\big)
\eeq
and the specific form \eqref{Hhatsummary} come from the explicit expressions of these local densities and currents.

The following result characterises $H_\psi$ in this language.
\begin{theorem}[generalised $T\b T$-deformation] \label{thsummaryttbar}
$H_\lambda := H_{\lambda\psi}$ satisfies \eqref{Hhatsummary}, with deformation function
\beq
	w(x,\theta) = \frc12 \psi_x(x,\theta).
\eeq
Thus $H_\psi$ is a generalised $T\b T$-deformation (Definition \ref{defigenTTbar}) of the system of free particles $H_0 = \sum_i \frc{p_i^2}2$.
\end{theorem}
Thus, the semiclassical Bethe systems form a complete family of generalised $T\b T$-deformations of the system of free particles on the line (we have not shown uniqueness of the solution to \eqref{Hhatsummary}, but we expect so). Combining with our discussion in Subsection \ref{ssectarbitrary}, our result gives credence to the claim made in \cite{10.21468/SciPostPhys.13.3.072} that {\em arbitrary two-body scattering shifts can be achieved from generalised $T\b T$-deformations.} 

\begin{rema}
The generalised $T\b T$-deformations introduced in \cite{10.21468/SciPostPhys.13.3.072} in fact correspond to a special choice of the $x$ dependence of the deformation function $w(x,\theta)$: it is $w(x,\theta) = \delta(x) w(\theta)$, involving the Dirac delta function $\delta(x)$ (as this choice is not a differentiable function, it is formally outside of our framework). In this case, with our requirements, $w(\theta)$ must be non-decreasing and anti-symmetric. Above, we have therefore extended the deformations to include more general spatial dependence. The further specialisation to deformations by products of the ``conventional" family of local conserved densities and currents, the original $T\b T$-deformations of \cite{smirnov_space_2017}, corresponds to the choice of $w(\theta)$ as a polynomial $w(\theta) = \sum_{c=1,3,4,\ldots} w_c\theta^c$ (only odd polynomials arise because of our anti-symmetry condition), and is expressed in terms of conventional local densities and currents as
\beq\label{TTbarconventional}
	\frc{\p H_{\lambda}}{\p\lambda}
	= \sum_{a,b} \int \dd x\, w_{ab}\big( q^\lambda_{a}(x)j^\lambda_{ b}(x) - j^\lambda_{ a}(x)q^\lambda_{ b}(x)\big)
\eeq
for weights $w_{ab} = w_{a+b} \lt({a+b\atop a}\rt)\in\R$. In particular, the mass-momentum deformation, showed in \cite{cardy2022t} to transform point particles into rods of finite lengths $-\lambda$, is $w(\theta) = \theta/2$.
\end{rema}

\subsection{Thermodynamic Bethe ansatz and local density approximation}

Finally, we look at the thermodynamics. We consider the specific free energy
\beq\label{fLsummary}
	f_L = -L^{-1} \log \sum_{N=0}^\infty
	\frc1{N!}\int \dd^Nx \dd^Np \,
	\exp\Big[-\sum_{i=1}^N\beta\big(x_i/L,\theta_i(\bs x,\bs p)\big)\Big].
\eeq
Here $L$ is the macroscopic scale, and $\beta(x,\theta)\in\R\cup\{\infty\}$ is a generalised chemical potential function. One may put sharp boundaries at $\pm L/2$, thus bounding a volume $L$ where the system lies, by choosing $\beta(x,\theta)=\infty$ for $|x|\geq 1/2$. A slowly-varying potential, which varies on large scales $L$, is obtained by choosing $\beta(x,\theta)$ to be, say, smooth in $x$. The specific free energy in infinite volumes is
\beq
	f = \lim_{L\to\infty} f_L.
\eeq
The form \eqref{fLsummary} is the grand-canonical partition function for the Boltzmann weight
\[
	\exp\Big[-\sum_{i=1}^N\beta\big(x_i/L,\theta_i(\bs x,\bs p)\big)\Big]
	=
	\exp\Big[-\int \dd x\dd\theta \,\beta\big(x/L,\theta\big)q_\theta(x)\Big]
\]
where the continuum of local densities $q_\theta(x)$ is involved.

In integrable models, one expects $f$ to be given by a combination of the local density approximation (LDA) and the TBA (if $\beta(x,\theta)$ is constant in $x$ except for a bounding volume, then no LDA is involved); and for $L$ finite, one does not expect any simple expression for $f_L$. For short we refer to ``local TBA" the combination of the LDA and TBA. As far as we are aware, the only rigorous results showing the local TBA for a large family of functions $\beta(x,\theta)$ in an interacting integrable system is that of \cite{percus1976,bulchandani2023modified} for the hard rods; this also gives the only known result for a TBA-like equation at $L$ finite.

In our family of models, we will show that {\em the specific free energy $f_L$ satisfies equations of TBA type}; this is expressed in Theorem \ref{theoTBA}. This is surprising, as it applies to the full generality of our model, hence of the two-body scattering shift. We believe this is a more general phenomenon in integrable systems.

Further, we will show that the local TBA indeed describes $f$. The requirements of the following theorem admit a wide range of possible generalised chemical potential functions, with or without sharp boundaries.
\begin{theorem}[Thermodynamic Bethe ansatz]\label{thsummaryTBA} Assume that $e^{-\beta(x,\theta)}$ is $x,\theta$-uniformly continuous in $x$ for a.e.~$(x,\theta)\in\R^2$, and that
\begin{equation}
\label{eq:integralphisummary}
	\int_\R \dd\theta\, \varphi(\theta) = \varphi_{\rm tot} < \infty,\quad
    \int_{\R^2} \dd x\dd \theta\,
    e^{-\beta(x,\theta)}
    = F < \infty,\quad
	\sup_{(x,\theta)\in\R^2} e^{-\beta(x,\theta)} =
	e^{-\beta_{-}} <\infty
\end{equation}
and
\begin{equation}\label{eq:boundCE}
	e^{-\beta_{-}}\varphi_{\rm tot} < \pi \big(\sqrt{1+4e^{-1}}-1\big) = 1.797324....
 < 2\pi e^{-1}.
\end{equation}
Then the free energy at infinite volume, $f$, exists and is given by
\begin{equation}\label{eq:fTBAsummaru}
    f=-\int_{\R^2} \frc{\dd x\dd \theta}{2\pi}\, e^{-\varepsilon(x,\theta)}
\end{equation}
where $\varepsilon(x,\theta)\in\R$ is the unique solution to
\begin{equation}\label{eq:varepTBAsummary}
    \varepsilon(x,\theta)=\beta(x,\theta)-\int_{\R}\frc{\dd\theta'}{2\pi}\,\varphi(\theta-\theta')e^{-\varepsilon(x,\theta')}
    \qquad \mbox{a.e. }x\in\R,\quad \mbox{all }\theta\in\R.
\end{equation}
Thus, the free energy takes the local thermodynamic Bethe ansatz form, with Maxwell-Boltzmann statistics and two-body scattering shift $\varphi(\theta-\theta')$ (Definition \ref{defiTBAlocal}). Further, we have the bounds
\beq\label{eq:boundsolutionsummary}
 -\frc{F}{2\pi}\frc{e^{-\beta_-}\varphi_{\rm tot}}{2\pi e^{-1} - e^{-\beta_-}\varphi_{\rm tot}}\leq f\leq 0,
 \qquad
e^{-\varep(x,\theta)}\varphi_{\rm tot} < 2\pi\quad
	\forall\;x,\theta\in\R.
\eeq
(See the full statements in Theorem \ref{theoTBA2}).
\end{theorem}

\begin{rema}
Note that the Boltzmann weight in \eqref{fLsummary} is the most general form of the generalised Gibbs ensembles (GGEs) \cite{vidmar2016generalized,essler2016quench,Ilievski_2016}. A sum of conventional local conserved densities in the exponent, $\sum_a \beta_a(x)q_a(x)$, as is usually taken for GGEs, is obtained again by taking $\beta(x,\theta) = \sum_a \beta_a(x)\theta^a$ as a polynomial in $\theta$.
\end{rema}

\section{Properties of classical Hamiltonian many-particle systems}\label{sectProperties}

In this section, we gather the basic definitions of the important concepts about classical many-body dynamics that are referred to in  Theorems \ref{thsummary}, \ref{thsummaryttbar} and \ref{thsummaryTBA}. These concepts -- Liouville integrability and finite-range interactions, elastic and factorised scattering, $T\b T$-deformations, and the thermodynamic Bethe ansatz -- are well known in many-particle integrable systems. However it is convenient here to formalise them, in a way that is physically sensible and at least broadly in agreement with previous investigations. This will make our main statements entirely unambiguous.

We use the notation $\I_n = \{1,2,\ldots,n\}$, as well as $\bs x = (x_i)_{i=1,\ldots,n}$ and $\bs x_{I} = (x_i)_{i\in I}$ for subsets $I$ (with the understanding that the ordering is in increasing order of the indices within $I$). We take the physical phase space to be $\R^{2N}$, with coordinates $(\bs x,\bs p)$ and canonical Poisson structure
\beq
	\{x_i,x_j\}=0,\quad \{p_i,p_j\}=0,\quad \{x_i,p_j\}= \delta_{ij}\qquad (i,j\in\I_N).
\eeq
A Hamiltonian $H\in C^1(\R^{2N})$ generates the dynamics in the usual way,
\beq
	\dot x_i = \{x_i,H(\bs x,\bs p)\},\quad \dot p_i = \{p_i,H(\bs x,\bs p)\}.
\eeq
The trajectories are fully determined once the initial conditions $(\bs x(0),\bs p(0))$ are fixed.

A {\em many-particle system} should be understood not as a single Hamiltonian $H$, but as a sequence of Hamiltonians, for $N=1,2,3,\ldots$. That is, the phase space is the direct sum of $\R^{2N}$ for $N=1,2,3,\ldots$. This is important in order to be able to take the thermodynamic limit $N\to\infty$ and consider aspects of locality, and will be implicitly understood below. For this sequence to represent a ``unique" many-particle system, interaction potentials in the sequence should be related to each other in appropriate ways. We discuss such relations in the next subsection.

In order to keep the general definitions as targeted as possible to the models of interest, {\bf we restrict ourselves to many-particle systems that are invariant under re-labelling of the particles}. That, is the Hamiltonian is assumed to have the symmetry
\beq\label{relabelling}
	H(\bs x,\bs p) = H(\bs x_\sigma,\bs p_\sigma)
\eeq
where $(\bs x_\sigma)_i = x_{\sigma(i)}$, etc., and $\sigma$ is any permutation of $N$ elements. This means that the interaction only depends on the positions and momenta of the particles, not on their labels. Most definitions are easily extended to situations where the particle labels matter.

\begin{rema}\label{remaphasespace}
In general, the phase space may be a subset $\mathcal S \subset \R^{2N}$, invariant under the dynamics. A typical nontrivial example, where particle labels matter, is the ``ordered" phase space, $\mathcal S = \{(\bs x,\bs p)\in\R^{2N}: x_{i+1}-x_i > a\;\forall\;i\}$ for some $a\in\R$, occurring with interactions that forbid particles to pass through each other. We keep $\mathcal S=\R^{2N}$ for simplicity, as this is the case needed here.
\end{rema}

\subsection{Liouville integrability and finite-range interactions}\label{ssectliouville}

Recall that a $N$-particle Hamiltonian $H$ is {\em Liouville integrable} (or Liouville-Arnold integrable) if there exists $N$ functionally independent non-constant differentiable functions (one often assumes them to be smooth) $Q_a(\bs x,\bs p)$, $a =1,2,\ldots,N$ on $\R^{2N}$ such that $\{Q_a,Q_b\}=0$ for all $a,b$, and $H = \sum_a c_a Q_a$ for some constants $c_a$ (not all zero). These are the conserved quantities.

For a many-particle system with unbounded trajectories, however, in order to have integrability one must impose additionally a short-range constraint on the interactions associated to the conserved quantities; here we restrict ourselves to a stronger finite-range constraint. In discussing the notion of interaction, there is an understanding that we separate the Hamiltonian into a non-interacting ``kinetic part'', defined via a dispersion relation $E\in C^1(\R)$ (independent of $N$) as $\sum_i E(p_i)$, plus a quasi-potential $V\in C^1(\R^{2N})$ (dependent on $N$; recall that it is ``quasi" as we admit dependence on momenta). The finite-range constraint we propose, with respect to a choice of dispersion relation, is a precise clustering condition for the quasi-potentials of each conserved quantity, that represents their finite-range interactions. We show how it implies good locality properties of conserved densities and currents. We will refer to this concept as {\em Liouville integrability with finite-range interaction}.

In order to express the clustering property of the quasi-potential, we need the following. For every $\ep>0$ and $\bs x\in\mathcal \R^N$, we define the unique partition $P_\ep(\bs x) = \{I_1,I_2,\ldots, I_n\}$ of $\I_N$ (i.e.~$I_r\cap I_s = \emptyset \;\forall\;r\neq s,\ \cup_r I_r = \I_N$) into ``clusters": the partition with the smallest $\sum_r |I_r|$ such that $|x_i-x_j|>\ep\;\forall\;i\in I_r,\;j\in I_s,\;r\neq s$. Jumping on the particles from the left-most to the right-most (say), one gathers them into clusters, a different cluster being made every time a distance larger than $\ep$ is jumped. Thus $\bs x_{I}$, for every $I\in P_\ep(\bs x)$, is a cluster of positions so that each nearest neighbours are a distance $\ep$ or less from each other, and that any two different clusters are a distance larger than $\ep$ from each other.
\begin{defi}[Finite-range Liouville integrable system]\label{defilocal}{\ }
\begin{enumerate}
\item A quasi-potential $V$ has {\em finite range} $\ep>0$ if it satisfies the following clustering condition:
For every $n\in \I_N$ there exists $V^{(n)}\in C^1(\R^{2n})$, such that:
\beq\label{condpotential}
	V(\bs x,\bs p) = \sum_{I\in P_\ep(\bs x)} V^{(|I|)}(\bs x_{I},\bs p_{I})\quad \forall\ (\bs x,\bs p)\in\R^{2N}.
\eeq
Further, $V^{(1)}(x,p)$ is a constant (which we set to $0$ without loss of generality).
\item A quantity $Q\in C^1(\R^{2N})$ has {\em interaction of finite-range} $\ep>0$ with respect to a dispersion relation $E(p)$ if it can be written in the form
\beq\label{Qalocal}
	Q(\bs x,\bs p) = \sum_i E(p_i) + V(\bs x,\bs p)
\eeq
for some quasi-potential $V(\bs x,\bs p)$ of finite range $\ep$.
\item The many-particle system with Hamiltonian $H$ (more precisely, with sequence of $H$'s for $N=1,2,\ldots$) is {\em Liouville integrable with finite-range interaction} if $H$ is Liouville integrable for every $N$, and if there are $E_a\in C^1(\R),\,a=1,2,\ldots$ (all independent of $N$) such that $\sum_{i=1}^N E_a(p_i),\,a=1,2,\ldots,N$ are functionally independent, and such that the conserved quantities
\begin{equation}
    Q_a(\bs x,\bs p) = \sum_i E_a(p_i) + V_a(\bs x,\bs p), \quad a=1,2,\ldots
\end{equation}
have uniformly (over $N$) finite-range interactions with respect to the $E_a$'s. That is, all $V_a$'s satisfy \eqref{condpotential} for a single $\ep>0$ and where the $V^{(n)}_a$'s do not depend on $N$.
\end{enumerate}
\end{defi}
Note that, with $N$ particles, the quantities $Q_a(\bs x,\bs p)$ are functionally independent for $a=1,2,\ldots,N$, as when coordinates are well separated, by clustering of the quasi-potential the resulting functions are $\sum_i E_a(p_i)$, which are functionally independent by assumption. Also, as we restrict to systems invariant under re-labelling, Eq.~\eqref{relabelling}, $Q_a$'s, hence $V_a^{(n)}$'s, are invariant.
 
Our finite-range definition means that if particles' positions cluster into separate groups, with different clusters separated by distances of more than $\ep$, then the interaction, for every conserved charge, separates into a sum of terms, one for each cluster. Thus, particles $i$ and $j$ do not interact with each other if they are a distance greater than $\ep$ from each other, and there is no chain of particles, each a distance $\ep$ or less to the next on the chain, that connects particles $i$ with $j$. Definition \ref{defilocal} makes it clear how a sequence of $H$, for $N\in\N$, consistently corresponds to a single many-particle system: the system is defined by a unique set of $E_a$ and of $V_a^{(n)}$ for $a,n\in\N$, and a unique $\ep$.

For simplicity, we concentrate from now on on kinetic parts with {\em polynomial form}: we assume that $E_a(p)= p^a$, that is
\beq\label{Qa}
	Q_a(\bs x,\bs p) = \sum_i p_i^a + V_a(\bs x,\bs p), \quad a=0,1,2,\ldots.
\eeq
This is well adapted to the Galilean dispersion relation, on which we focus; it is a simple matter to adapt the discussion in this paper to other functions $E_a(p)$. Note that we added for convenience the index $a=0$, with $V_0=0$ and $Q_0=N$. The Hamiltonian is taken as
\beq\label{HQ2}
    H = \frc12 Q_2 =   \sum_i \frc{p_i^2}2 + V(\bs x,\bs p)
\eeq
where $V = \frc12 V_2$.

\begin{rema}
In fact, the condition \eqref{condpotential} says a bit more than the interaction being finite-range: it says that the particle labelling is not important (as all restricted quasi-potentials are the same) and that there is no ``external force" acting on single particles (as $V^{(1)}(x,p)$ is a constant). These are for simplicity, and clearly not essential for the interaction between particles to be considered finite-range. One may also extend this definition to short- / long-range interactions, where corrections to quasi-potential clustering are exponentially / algebraically decaying with the distances between the clusters $I_r$. The above is sufficient for our purposes.
\end{rema}

\subsubsection{Conserved densities and currents}

In order to understand the meaning of finite-range interactions, and physically justify our definition, it is helpful to see how it allows us to construct a {\em charge density and a current} with good locality properties, for a charge $Q_a(\bs x,\bs p)$. Charge densities and currents will play an important role in the description of $T\b T$-deformations.

A charge density for the conserved quantity $Q_a(\bs x,\bs p)$ is a distribution-valued phase-space function, which we denote by $q_a(\bs x,\bs p;x)$, such that
\beq\label{integratedq}
	\int \dd x\,q_a(\bs x,\bs p;x) = Q_a(\bs x,\bs p).
\eeq
It is required that $q_a$ be a $C^1$ distribution-valued function, by which we mean that for any Schwartz function $f:\R\to\R$, the resulting $(\bs x,\bs p)\mapsto \int \dd x\,q_a(\bs x,\bs p;x)f(x)$ is in $C^1(\R^{2N})$. A current associated to the charge density is also a distribution-valued phase-space function, such that the following conservation law
\beq\label{cont}
	\{q_a(\bs x,\bs p;x),H\} + \p_x j_a(\bs x,\bs p;x) = 0
\eeq
holds in the distributional sense -- integrated against any Schwartz function.

If $Q_a(\bs x,\bs p)$ has finite-range interaction, then one may define $q_a(\bs x,\bs p;x)$ and $j_a(\bs x,\bs p;x)$ as follows. For every $n$, choose quasi-potentials $V^{(n,r)}\in C^1(\R^{2n}),\;r=1,\ldots,n$ such that $V^{(n)} = \sum_{r=1}^n V^{(n,r)}$. This should be interpreted as a separation of the quasi-potential $V^{(n)}(\bs x_{\I_n},\bs p_{\I_n})$ into the individual contributions of the particles $r=1,2,\ldots,n$; but here it is just a general construction. With re-labelling invariance \eqref{relabelling}, one may additionally ask
\beq\label{Vnrrelabl}
	V^{(n,\sigma(r))}(\bs x_{\I_n},\bs p_{\I_n}) = V^{(n,r)}((\bs x_{\I_n})_\sigma,(\bs p_{\I_n})_\sigma)
\eeq
for all $n$-element permutations $\sigma$, although this is not required for the general construction. In constructing conserved densities for tracer dynamics below, we will make a particular physically sensible choice which satisfies \eqref{Vnrrelabl}.

Then, given such separations of the quasi-potentials into individual particle contributions, we define
\beq\label{defqgeneral}
	q_a(\bs x,\bs p;x) := \sum_{i=1}^N\delta(x-x_i)\Big(
	p_i^a + \sum_{I\in P_\ep(\bs x)}\sum_{r=1}^{|I|}V^{(|I|,r)}(\bs x_I,\bs p_I)\delta_{i,I(r)}\Big)
\eeq
where $I(r)$ is the $r^{\rm th}$ element of the index subset $I$ (in increasing order of indices). This reproduces \eqref{integratedq}. Note that for every $\bs x,\bs p$ (and every $N$), the resulting $q_a(\bs x,\bs p;x)$ is a finite sum of Dirac delta functions supported on $x_i$'s. Further, writing
\[
	A_{a,i}(\bs x,\bs p) = p_i^a + \sum_{I\in P_\ep(\bs x)}\sum_{r=1}^{|I|}V^{(|I|,r)}(\bs x_I,\bs p_I)\delta_{i,I(r)},
\]
we have $q_a(\bs x,\bs p;x) = \sum_{i=1}^N\delta(x-x_i) A_{a,i}(\bs x,\bs p)$ and the current can be defined as
\beq\label{defjgeneral}
	j_a(\bs x,\bs p;x) := \sum_{i=1}^N\Big(\dot x_i(\bs x,\bs p)\delta(x-x_i)A_{a,i}(\bs x,\bs p) - \Theta(x-x_i) \dot A_{a,i}(\bs x,\bs p)\Big)
\eeq
where $\Theta$ is Heaviside's step function and $\dot A_{a,i}(\bs x,\bs p)= \{A_{a,i}(\bs x,\bs p),H\}$. This satisfies the continuity relation \eqref{cont}.

The important aspect of this (somewhat formal) construction is that both the densities and currents have ``good" locality properties. In order to express these, we define
\beq
	\ell_\ep:\R^{N}\to \R^+\ :\  \ell_\ep(\bs x)
	= \max(\ep |I|\,:\, I \in P_\ep(\bs x)).
\eeq
This is the largest distance that can separate two positions that pertain to the same cluster, over all available clusters. Thus $\ell_\ep(\bs x)$ is the {\em effective interaction range} of the configuration $\bs x$, given an interaction of finite range $\ep$: it is the maximal distance over which there can be interaction between two particles.
\begin{lemma}[Finite-range properties of densities and currents]\label{lemlocality} In the distributional sense,
\beq
	\{q_a(\bs x,\bs p;x),q_b(\bs x,\bs p;x')\}= 
	\{q_a(\bs x,\bs p;x),j_b(\bs x,\bs p;x')\}=
	\{j_a(\bs x,\bs p;x),j_b(\bs x,\bs p;x')\}=0
\eeq
for every $a,b$, and every $x,x'\in\R$ such that $|x-x'|>4\ell_\ep(\bs x)$. For the equations involving currents to make sense, we require additionally that $j_a$ be a $C^1$ distribution-valued function; this holds in particular if all quasi-potentials are in $C^2(\R^{2N})$.
\end{lemma}
\proof What we show is that the conserved densities $q_a(\bs x,\bs p;x)$ do not depend on $x_i, p_i$ for all $i, x$ such that $|x_i-x|>\ell_\ep(\bs x)$, and the conserved currents $j_a(\bs x,\bs p;x)$ do not depend on $x_i, p_i$ for any $i$ such that $|x_i-x|>2\ell_\ep(\bs x)$. For the densities, the statement is clear from \eqref{defqgeneral}. For the currents, because of the finite-range form of $H$, the first term inside the sum in \eqref{defjgeneral} depends only on particles (that is, on their positions and momenta) whose positions are within a distance $\ell_\ep(\bs x)$ of $x$. Likewise, for the second term, the quantity $\dot A_{a,i}(\bs x,\bs p)$ depends only on particles whose positions are within a distance $2\ell_\ep(\bs x)$ from $x_i$. As $\sum_i \dot A_{a,i}(\bs x,\bs p) = \dot Q_a = 0$, then $\sum_{i:x_i\leq x} \dot A_{a,i}(\bs x,\bs p) = - \sum_{i:x_i > x} \dot A_{a,i}(\bs x,\bs p)$. By using either the left or right-hand side of this equality, one concludes that $\sum_{i:x_i\leq x} \dot A_{a,i}(\bs x,\bs p)$ must depend only on particles whose positions are within a distance $2\ell_\ep(\bs x)$ from $x$.
\eproof

The above allows us to consider the system for infinitely-many particles. A configuration for infinitely-many particles is ``good" if $|\{I\in P_\ep(\bs x):I\subset[-x,x]\}|$ increases unboundedly as $x\to\infty$ (that is, the number of pairs of neighbouring particles a distance larger than $\ep$ from each other, lying within $[-x,x]$, increases unboundedly with $x$). For many-particle systems that are Liouville integrable with finite-range interactions, densities and currents of conserved quantities Poisson-commute at large enough distances in any good configuration, and time evolution -- at least for small enough times -- is well defined. One would expect good configurations to stay good in time, and to arise almost surely in thermodynamic states. This justifies our definition of finite-range interactions.

\subsection{Elastic, factorised scattering and tracer dynamics}\label{ssectelastic}

Another important property of many-particle integrable systems  on the line is that scattering processes are elastic and factorised. We provide here a precise definition of this property, in the simplest cases. The elastic, factorised scattering condition below is a formalisation of a standard concept, but the property of the dynamics being a {\em tracer dynamics} is something which, to our knowledge, has not been fully appreciated before.

Recall that $\R^{2N}_{\neq}:=\{(\bs y,\bs \theta)\in\R^{2N}:\theta_i\neq \theta_j\;\forall\; i\neq j\}$. Consider a function $\varphi\in C^1(\R^2)$.
\begin{defi}\label{defiscattering}
We say that the many-body system $H$ satisfies the {\em elastic, factorised scattering condition}, with two-body scattering shift $\varphi(\theta,\theta')$, if for every $N$, there exists a dense subspace $\mathcal S'\subset \R^{2N}$ invariant under time evolution, such that for every initial condition $(\bs x(0),\,\bs p(0))\in\mathcal S'$: (1) the trajectories become linear with non-coinciding asymptotic momenta $\theta_i\in\R$, $\theta_i\neq\theta_j$ for $i\neq j$ and ``impact parameters" $x_i^\pm\in\R$ at large times,
\beq\label{asymptotic}
	p_i(t) = \theta_{\sigma^\pm (i)} + o(1), \qquad
	x_i(t) = \theta_{\sigma^\pm (i)} t + x_{\sigma^\pm(i)}^\pm + o(1)
	\qquad (t\to\pm\infty)
\eeq
where $\sigma^-(i) = i$ (this just defines the $\theta_i$'s) and $\sigma^+ =: \sigma$ is a permutation of $N$ elements; (2) the impact parameters are related by a sum of individual two-body shifts,
\beq\label{factorised}
	x_{i}^+-x_{i}^-
	= \sum_{j\neq i} \sgn(\theta_j-\theta_i)
	\varphi(\theta_i,\theta_j);
\eeq
and (3) the map $\mathcal S'\to \mathcal \R^{2N}_{\neq}: (\bs x(0),\,\bs p(0)) \mapsto (\bs x^-,\bs \theta)$ is a bijection.  We say, in addition, that $H$ is a {\em tracer dynamics} if for every $N$, $\sigma = \1$ (the identity permutation) for all $(\bs x(0),\,\bs p(0))\in\mathcal S'$.
\end{defi}

Here, elastic scattering is the statement that it is the same set of momenta $\{\bs\theta\}$ that occurs both at negative and positive infinite times -- thus a permutation $\sigma$ exists. Factorised scattering is the statement \eqref{factorised}: for the case $N=2$, this fixes $\varphi(\theta_1,\theta_2)$ as the shift incurred by particle 1 upon scattering with particle 2, and for $N>2$ this indicates that the total shift of particle $i$ is obtained by summing over the two-body shifts that would be incurred by independent two-body collisions with other particles. In the case of a tracer dynamics, each trajectory $x_i(t)$ is associated with a unique asymptotic momentum $\theta_i$ -- thus the trajectory is a ``tracer" of this asymptotic momentum, and effectively carries the information of the asymptotic momentum even within configurations where particles are not well separated and their momenta are very different from the asymptotic momenta.

Note that the bijection condition says in particular that the asymptotic parameters $(\bs x^-,\bs \theta)$ can be chosen arbitrarily in $\R^{2N}_{\neq}$. These can be seen as alternative initial conditions, which fix the scattering problem by fixing the asymptotic trajectories as $t\to-\infty$. Then, $\sigma$ and $\bs x^+$ are determined by the dynamics.

The coordinates $\bs x^-,\bs \theta$ are canonical. Indeed, the canonical Poisson bracket is preserved by time evolution, thus we just have to look at the asymptotic expansion at large times of $\{x_i(t),x_j(t)\}=0,\ \{p_i(t),p_j(t)\}=0$ and $\{x_i(t),p_j(t)\}=\delta_{i,j}$. Further, they evolve trivially under the dynamics:
\beq
	\theta_i(t)= \theta_i(0),\quad x^-_i(t) = \theta_i t + x^-_i(0)
\eeq
as is immediate from \eqref{asymptotic}.

Note that if $\varphi(\theta,\theta')>0$ then the scattering shifts correspond to {\em positive time delays} at head-on collisions. In particular, in the geometric interpretation of scattering (see Subsection \ref{ssectarbitrary} and discussions in \cite{Doyon2018,DHY23b}), this corresponds to an addition of effective space by the amount $\varphi(\theta,\theta')$, as perceived by a particle with asymptotic momentum $\theta$, in the presence of a particle with asymptotic momentum $\theta'$.

\begin{rema}
Following Remark \ref{remaphasespace}, one may extend this definition to phase spaces $\mathcal S\subset \R^{2N}$, with $\mathcal S'\subset \mathcal S$ a dense subset. Also, in certain problems it may be more appropriate to consider different asymptotic in-spaces $\mathcal S^- \subset \R^{2N}_{\neq}$, with $\mathcal S'\to \mathcal S^-$ being a bijection. In any case, the asymptotic in-space must satisfy $\mathcal S^- = \{(\bs y+ t\bs\theta,\bs \theta): (\bs y,\bs \theta) \in \mathcal S^-,\,t\in\R\}$.
\end{rema}

\begin{rema}
It is of course simple to extend to many particle species. Further, the above defines an ``asymptotically free" elastic, factorised scattering dynamics, as we do not allow bound states. In dynamics where bound states can form, the bijection condition would be broken, and one would have to extend the space of asymptotic particles to include bound states. Alternatively, an appropriate choice of phase space $\mathcal S$ may be possible to avoid the formation of bound state. Naturally, scattering is asymptotically free if the interaction between particles is repulsive. For simplicity we concentrate on asymptotically free scattering.
\end{rema}

\subsubsection{Finite-range tracer dynamics: conserved densities for asymptotic momenta}

One expects a close relation between elastic, factorised scattering and finite-range Liouville integrability. Let us discuss briefly three relations; the third is of most interest here.

{\em From finite-range Liouville integrability to elastic, factorised scattering.} Finite-range Liouville integrability can be expected to guarantee elastic, factorised scattering. This is because of the argument \cite{parke_absence_1980,doyon_lecture_2019} that conserved charges $Q_a$ can be used to shift trajectories in such a way that particles only interact in well separated two-body processes. It should be straightforward to make this rigorous, assuming some basic description of asymptotic states; this is beyond the scope of this paper.

{\em Finite-range Liouville integrability and elastic, factorised scattering: charges in terms of asymptotic momenta.} If $H$ satisfies the elastic, factorised scattering condition, then the following distribution-valued functions of phase space (for distributions in $\theta\in\R$) are independent of time:
\beq\label{Qtheta}
	Q_\theta(\bs x,\bs p) = \sum_i \delta(\theta-\theta_i(\bs x,\bs p)).
\eeq
In particular, one may construct the Liouville conserved quantities as
\beq\label{Qatheta}
	Q_a(\bs x,\bs p) = \int \dd\theta\, \theta^a Q_\theta(\bs x,\bs p)  = \sum_i \theta_i(\bs x,\bs p)^a.
\eeq
If $H$ is in addition Liouville integrable with finite-range interaction, then the $Q_a$'s in the form \eqref{Qa} agree with \eqref{Qatheta} at least on the dense subspace $\mathcal S'\subset \R^{2N}$. Indeed, they are constant in time, so we may evaluate them on $(\bs x,\bs p)\in \mathcal S'$ evolved to any finite but large enough time; the scattering condition \eqref{asymptotic} and the finite-range property of the quasi-potentials (Definition \ref{defilocal}) imply that they take the form \eqref{Qatheta}. This implies strong short-range constraints on $\sum_i \theta_i(\bs x,\bs p)^a$. Combining with the previous discussion, we see that short-range Liouville integrability along with some basic description of asymptotic states should give much structure to the scattering theory; again this is beyond the scope of this work.

{\em From elastic, factorised scattering to finite-range Liouville integrability.} Can we go in the opposite direction? Assuming $H$ satisfies the elastic, factorised scattering condition, and perhaps that it has a finite-range potential, can we deduce finite-range Liouville integrability? Do the $Q_a$'s defined in \eqref{Qatheta} have finite-range interactions? These are interesting questions that we cannot address here. However, with an additional condition on the scattering, {\em finite-range tracer dynamics}, we can indeed answer positively, as we now explain.
\begin{defi}\label{defifinitetracer}
A many-particle system $H$ that satisfies the elastic, factorised scattering condition and that is a tracer dynamics (Definition \ref{defiscattering}) is a {\em finite-range tracer dynamics} with interaction range $\ep>0$, if for every $n\geq 0$ and $r=0,1,\ldots,n$ there is $\theta^{(n,r)}(\bs x_{\I_n},\bs p_{\I_n})$ in $C^1(\R^{2n})$, such that for every $(\bs x,\bs p)\in\R^{2N}$, $I\in P_\ep(\bs x)$ and $r=0,1,\ldots,|I|$, we have $\theta_{I(r)}(\bs x,\bs p) = \theta^{(|I|,r)}(\bs x_{I},\bs p_{I})$.
\end{defi}
Recall that $I(r)$ is the $r^{\rm th}$ element of the index subset $I$ (in increasing order of the indices), and that $P_\ep(\bs x)$ is the partition into clusters of interacting particles, defined in the paragraph above Definition \ref{defilocal}.

The definition means that the information of the asymptotic momentum for particle $i$ is already included within the interacting cluster to which this particle belongs -- there is no need for the information about other particles (with which, of course, particle $i$ will nevertheless interact in passing from time 0 to asymptotic times). Further, the information of the label of the particle is also irrelevant. In fact, as we restrict to systems invariant under re-labelling, Eq.~\eqref{relabelling}, we have the re-labelling invariance
\beq
	\theta^{(n,\sigma(r))}(\bs x_{\I_n},\bs p_{\I_n}) = \theta^{(n,r)}((\bs x_{\I_n})_\sigma,(\bs p_{\I_n})_\sigma).
\eeq

If $H$ is a finite-range tracer dynamics, then it is easy to show that it is Liouville integrable with finite-range interaction. Indeed, one can choose, in Definition \ref{defilocal},
\beq
	V^{(|I|)}_a(\bs x_{I},\bs p_{I}) = \sum_{i\in I} (\theta_i^a(\bs x,\bs p) - p_i^a)
\eeq
and all charges \eqref{Qatheta} now have the form \eqref{Qa}, with finite-range interactions. One can also construct the conserved densities and currents (see \eqref{defqgeneral}, \eqref{defjgeneral}). Making the choice $V_a^{(n,r)}(\bs x_{\I_n},\bs p_{\I_n}) = \big(\theta^{(n,r)}(\bs x_{\I_n},\bs p_{\I_n})\big)^a - p_r^a$ which satisfies \eqref{Vnrrelabl}, we have
\beq\label{qatheta}
	q_a(\bs x,\bs p;x) = \sum_{i=1}^N \delta(x-x_i)\theta_i^a(\bs x,\bs p),\quad
	j_a(\bs x,\bs p;x) = \sum_{i=1}^N \dot x_i(\bs x,\bs p) \delta(x-x_i)\theta_i^a(\bs x,\bs p);
\eeq
these have the locality properties stated at the end of Subsection \ref{ssectliouville}. Of course, one can construct a ``continuum" of conserved densities and currents: these are functions of phase space valued in distributions over $(x,\theta)\in\R^2$, defined as
\beq\label{qtheta}
	q_\theta(\bs x,\bs p;x) = \sum_i \delta(x-x_i)\delta(\theta-\theta_i(\bs x,\bs p)),\quad
	j_\theta(\bs x,\bs p;x) = \sum_i \dot x_i(\bs x,\bs p)\delta(x-x_i)\delta(\theta-\theta_i(\bs x,\bs p)).
\eeq
They satisfy $\{q_\theta(\bs x,\bs p;x),H\}+\p_x j_\theta(\bs x,\bs p;x) = 0$ in the distributional sense, and we recover $q_a(\bs x,\bs p;x) = \int \dd\theta\,\theta^a q_\theta(\bs x,\bs p;x)$, etc. Further, by Lemma \ref{lemlocality},
\beq
	\{q_\theta(\bs x,\bs p;x),q_{\theta'}(\bs x,\bs p;x')\}= 
	\{q_\theta(\bs x,\bs p;x),j_{\theta'}(\bs x,\bs p;x')\}=
	\{j_\theta(\bs x,\bs p;x),j_{\theta'}(\bs x,\bs p;x')\}=0
\eeq
for $|x-x'|>4\ell_\ep(\bs x)$, again in the distributional sense and with the additional differentiability properties for the currents (as in Lemma \ref{lemlocality}).

\subsection{Generalised $T\b T$-deformations}\label{ssectTTbar}

We discussed $T\b T$-deformations in Subsection \ref{ssectmainTTbar}. Here we provide the formal definitions.

It is convenient to start with a formalisation of the standard $T\b T$-deformations, for arbitrary local densities and currents. This only requires finite-range Liouville integrability (in fact, only at least one pair of local conserved density and current is required to exist):
\begin{defi}\label{defiTTbar}
For every $\lambda\in [0,1]$, let the many-body system $H_\lambda$ be Liouville integrable with finite-range interaction, and assume that $H_\lambda(\bs x,\bs p)$ is differentiable in $\lambda$ for all $(\bs x,\bs p)\in \R^{2N}$ (for every $N$). We say that $H_1$ is a {\em $T\b T$-deformation of $H_0$} if there exists finitely many functions $w_{a,b}\in C^1(\R)$ such that for every $N$,  $(\bs x,\bs p)\in\R^{2N}$ and $\lambda\in[0,1]$,
\beq\label{TTbarstandard}
	\frc{\p H_\lambda(\bs x,\bs p)}{\p \lambda}
	= \sum_{a,b} \int \dd y \dd z\, w_{a,b}(y-z)\big( q^\lambda_{a}(\bs x,\bs p;y)j^\lambda_{b}(\bs x,\bs p;z) - j^\lambda_{a}(\bs x,\bs p;y)q^\lambda_{b}(\bs x,\bs p;z)\big).
\eeq
\end{defi}
Here and below, we use the upper index $\lambda$ in order to represent phase-space functions that are constructed from the Hamiltonian $H_\lambda$. We note that the integrals on $y$ and $z$ can be evaluated as explicit phase space functions using the definitions \eqref{defqgeneral}, \eqref{defjgeneral}. Eq.~\eqref{TTbarstandard} defines a flow of Hamiltonians $[0,1]\ni\lambda\mapsto H_\lambda$, the $T\b T$-deformation of $H_0$ to $H_1$. In particular, the right-hand side of \eqref{TTbarstandard} is in general a non-linear functional of $H_\lambda$.

More interesting is the generalised $T\b T$ deformation proposed in \cite{10.21468/SciPostPhys.13.3.072}. We now have the right framework in order to define it rigorously: that of finite-range tracer dynamics.
\begin{defi}\label{defigenTTbar}
In the context of Definition \ref{defiTTbar}, let each many-body system $H_\lambda$ satisfy the elastic, factorised scattering property, and be a finite-range tracer dynamics, Definitions \ref{defiscattering} and \ref{defifinitetracer}. Let $\mathcal S_\lambda'\subset \R^{2N}$ be the dense subsets of $\R^{2N}$ involved in these definitions. Assume that $H_\lambda(\bs x,\bs p)$ is differentiable in $\lambda$ for all $(\bs x,\bs p)\in \R^{2N}$ (and every $N$). We say that $H_1$ is a {\em generalised $T\b T$-deformation of $H_0$} if there exists a function $w\in C^1(\R^2)$ such that  for every $N$,  $(\bs x,\bs p)\in \mathcal S_\lambda'$ and $\lambda\in[0,1]$,
\beqa
	\frc{\p H_\lambda(\bs x,\bs p)}{\p \lambda} &=& \int \dd\theta\dd\alpha
	\dd y\dd z \, w(y-z,\theta-\alpha)\big( q^\lambda_{\theta}(\bs x,\bs p;y)j^\lambda_{\alpha}(\bs x,\bs p;z)
	-j^\lambda_{\theta}(\bs x,\bs p;y)q^\lambda_{\alpha}(\bs x,\bs p;z)\big)\n
	&=& -\sum_{ij}
	\big(\dot x^\lambda_{i}(\bs x,\bs p) - \dot x^\lambda_{j}(\bs x,\bs p)\big)\,w\big(x_{i}-x_{j},\theta^\lambda_{i}(\bs x,\bs p)-\theta^\lambda_{j}(\bs x,\bs p)\big).
 \label{Hhat}
\eeqa
\end{defi}
Here we have explicitly evaluated the integrals in the generalised $T\b T$-deformation as a function on phase space, as densities and currents take a simple form for finite-range tracer dynamics, Eq.~\eqref{qtheta}. Note how, according to our definitions, the deformation equation \eqref{Hhat} makes sense on $\mathcal S'_\lambda$, but as this is a dense subset of $\R^{2N}$, this is a strong enough constraint on the family of Hamiltonians.

\begin{rema} In relation to Remark \ref{remaphasespace}, in general we may need to consider a $\lambda$-dependent physical phase space  $\mathcal S_\lambda\subset \R^{2N}$. For simplicity we assume that the physical phase space is not affected by the $T\b T$-deformation, which is sufficient for our models, but may not always be the case more generally.
\end{rema}


\subsection{Thermodynamic Bethe ansatz}
\label{ssectTBA}

In order to understand the thermodynamics of a system of particles, one may evaluate the specific free energy (per unit volume). In integrable systems, where one considers the generalised free energy, involving all conserved densities, this takes a very special form. The form goes by the name of thermodynamic Bethe ansatz (TBA) -- despite the term ``Bethe ansatz" in it, which reminds one of quantum models, this form is valid more generally (see e.g.~\cite{doyon_lecture_2019}). 

For a Liouville integrable system of with finite-range interactions, the generalised partition function in the finite volume $[-L/2,L/2]$ is defined by
\begin{equation}\label{eq:gge}
    Z_L=\sum_{N=0}^\infty\frac{1}{(2\pi)^NN!}\int_{[-L/2,L/2]^N\times\mathbb{R}^N}\prod_{j=1}^N\dd x_j\dd p_j\,e^{-\sum_{a=0}^b \beta_aQ_a(\bs x,\bs p)}.
\end{equation}
$Q_a(\bs x,\bs p)$, $a=0,1,2,\ldots$ are defined in \eqref{Qa} and $\beta_a\in\R$ are chemical potentials for the conserved charges $Q_a(\bs x,\bs p)$. Note how the factor $1/N!$ is natural under our assumption that the charges are invariant under re-labelling. As the integrands and each term in the series are non-negative, $Z_L$ always converges absolutely, but we will require that it converges to a finite value. The finite-volume specific free energy is
\begin{equation}\label{eq:freeenergyfinite}
    f_L=-L^{-1}\log Z_L
\end{equation}
and the specific free energy is
\begin{equation}
\label{eq:freeenergy}
     f = \lim_{L\to\infty} f_L
\end{equation}
if the limit exists.

We now formalise what it means for the specific free energy to take the TBA form. 
\begin{defi}\label{defiTBA}
Let a many-body system be Liouville integrable with finite-range interaction, Definition \ref{defilocal}, and satisfy the elastic, factorised scattering property, Definition \ref{defiscattering}. The specific free energy associated to \eqref{eq:gge}, i.e.~Eq.~\eqref{eq:freeenergy} with \eqref{eq:freeenergyfinite}, is said to take the TBA form for the free energy function $\mathsf F(\varep)$ and the two-body scattering shift $\varphi(\theta,\theta')$, if the partition function \eqref{eq:gge} converges for all $L>0$, and the limit \eqref{eq:freeenergy} exists and is given by
\begin{equation}\label{fTBA}
    f=\int_\mathbb{R}\frac{\dd \theta}{2\pi}\, \mathsf{F}(\varepsilon(\theta)),
\end{equation}
where the pseudo-energy $\varepsilon(\theta)$ solves the integral equation
\begin{equation}\label{pseudoenergydef}
    \varepsilon(\theta)=\sum_{a=0}^b \beta_a\theta^a+\int_\mathbb{R}\frc{\dd\theta'}{2\pi}\,\varphi(\theta',\theta)\mathsf{F}(\varepsilon(\theta')),
    \qquad \mbox{a.e. }\theta\in\R.
\end{equation}
\end{defi}
The free energy function may take different forms depending on the particle statistics of the system \cite{doyon_lecture_2019}. For classical particles, Eq.~\eqref{eq:gge}, it is natural to expect $\mathsf{F}(\varepsilon)=-e^{-\varepsilon}$, but this has to be proven. One may consider a more general definition where, in \eqref{fTBA}, the measure is $\dd p(\theta)$ for a given momentum function $p(\theta)$ (for instance, in relativistic models one takes $p(\theta) = \sinh(\theta)$); here, in agreement with our setup, the momentum function is simply taken as $p(\theta)=\theta$.

The interpretation of \eqref{pseudoenergydef} is clear: the specific free energy takes the form of that for free particles with contribution from particle $\theta$ to the ``energy" (negative logarithm of the Boltzmann weight) given by $\varep(\theta)$. This contribution is the bare ``energy", modified because the presence of a particle $\theta$ effectively changes, in the geometric interpretation of scattering, the volume perceived by all particles $\theta'$; the free energy gained by this change of volume must be added.

In this definition, we do not specify if the solution to \eqref{pseudoenergydef} is unique. Implied in the definition is that both $\mathsf F(\varepsilon(\theta))$ and $\varphi(\theta',\theta)\mathsf F(\varepsilon(\theta'))$ (for all $\theta,\theta'\in\R$) are Lebesgue measurable with finite integrals, and that the set of measure zero where \eqref{pseudoenergydef} might not hold, does not contribute to the integrals.

Of course, in Eq.~\eqref{eq:gge}, one may in principle put infinitely many charges, $b=\infty$. This brings the problem of convergence of the sum, which we will not discuss here in the level of generality of Definition \ref{defiTBA}. Instead, we extend Definition \ref{defiTBA} in the following three natural directions.

\medskip
{\bf Extension: tracer dynamics.} First, if the system is a finite-range tracer dynamics, Definition \ref{defifinitetracer}, then one may consider more generally the partition function
\begin{equation}\label{eq:ggetracer}
    Z_L=\sum_{N=0}^\infty\frac{1}{(2\pi)^NN!}\int_{[-L/2,L/2]^N\times\mathbb{R}^N}\prod_{j=1}^N\dd x_j\dd p_j\,e^{-\int \dd\theta\, \beta(\theta)Q_\theta(\bs x,\bs p)}.
\end{equation}
Here $Q_\theta(\bs x,\bs p)$ is defined in \eqref{Qtheta}. As $\int \dd\theta\, \beta(\theta)Q_\theta(\bs x,\bs p) = \sum_i \beta(\theta_i(\bs x,\bs p))$, it is sufficient for $e^{-\beta(\theta)}\in [0,\infty)$ to be Lebesgue measurable on $\theta\in\R$ for the integrals to make sense. Note that we admit positive infinite values for $\beta(\theta)$.
\begin{defi}\label{defiTBAtracer} In the setup of Definition \ref{defiTBA}, assume that the system is, in addition, a finite-range tracer dynamics, Definition \ref{defifinitetracer}. Paralleling Definition \ref{defiTBA}, the specific free energy associated to \eqref{eq:ggetracer} is said to take the TBA form if
\begin{equation}\label{fTBA_2}
    f=\int_\mathbb{R}\frac{\dd \theta}{2\pi}\, \mathsf{F}(\varepsilon(\theta)),
\end{equation}
where the pseudo-energy $\varepsilon(\theta)$ solves the integral equation
\begin{equation}\label{pseudoenergydeftracer}
    \varepsilon(\theta)=\beta(\theta)+\int_\mathbb{R}\frc{\dd\theta'}{2\pi}\,\varphi(\theta',\theta)\mathsf{F}(\varepsilon(\theta')),
    \qquad \mbox{a.e. }\theta\in\R.
\end{equation}
\end{defi}
For convergence to finite values, typically $\beta(\theta)$ must increase in $|\theta|$ sufficiently fast; although the precise constraints depend on the model. The case \eqref{eq:gge} is recovered by choosing $\beta(\theta) = \sum_{a=0}^b \beta_a\theta^a$. The partition function \eqref{eq:ggetracer} is the more accurate definition of the case $b=\infty$: there is no need for $\beta(\theta)$ to have a Taylor series expansion.

\medskip
{\bf Extension: local density approximation.} Second, of much interest is to take the chemical potentials to be space-varying. We consider
\begin{equation}\label{eq:ggelda}
    Z_L=\sum_{N=0}^\infty\frac{1}{(2\pi)^NN!}\int_{\mathbb R^{2N}}\prod_{j=1}^N\dd x_j\dd p_j\,e^{-\sum_{a=0}^b \int \dd x\,\beta_a(x/L)q_a(\bs x,\bs p;x)}
\end{equation}
or, when combined with the extension to finite-range tracer dynamics,
\begin{equation}
\label{eq:ggetracerlda}
    Z_L=\sum_{N=0}^\infty\frac{1}{(2\pi)^NN!}\int_{\mathbb R^{2N}}\prod_{j=1}^N\dd x_j\dd p_j\,e^{-\int \dd x \dd\theta\, \beta(x/L,\theta)q_\theta(\bs x,\bs p;x)}.
\end{equation}
Here, $q_a(\bs x,\bs p;x)$ and $q_\theta(\bs x,\bs p;x)$ are defined in \eqref{defqgeneral} and \eqref{qtheta}, respectively, and we have
\[
	\int \dd x \dd\theta\, \beta(x/L,\theta)q_\theta(\bs x,\bs p;x) = \sum_i \beta(x_i/L,\theta_i(\bs x,\bs p)).
\]
For the integrals to make sense, it is sufficient for $\beta_a(x)$'s to be measurable on $x\in\R$, or $\beta(x,\theta)$ to be measurable on $(x,\theta)\in\R^2$; again these functions take values in $\R\cup\{+\infty\}$. Note that it is possible to implement a sharp volume $L$ where the system lies within this formulation, simply by setting $\beta_a(x)$ for appropriate $a$ (corresponding to a strictly positive $q_a$), or $\beta(x,\theta)$ for all $\theta$, to be $+\infty$ for all $|x|\geq 1/2$. Thus, this is the most general formulation.
\begin{defi}\label{defiTBAlocal} Consider either the setup of Definition \ref{defiTBA} with partition function \eqref{eq:ggelda}, or of Definition \ref{defiTBAtracer} with partition function \eqref{eq:ggetracerlda}. For the former, set $\beta(x,\theta) = \sum_{a=0}^b\beta_a(x)\theta^a$. Paralleling these definitions, the associated specific free energy is said to take the local TBA form if
\begin{equation}\label{fTBAlocal}
    f=\int_{\mathbb{R}^2}\frac{\dd x\dd \theta}{2\pi}\, \mathsf{F}(\varepsilon(x,\theta))
\end{equation}
where the pseudo-energy $\varepsilon(x,\theta)$ solves the integral equation
\begin{equation}\label{pseudoenergydeflocal}
    \varepsilon(x,\theta)= \beta(x,\theta)+\int_\mathbb{R}\frc{\dd\theta'}{2\pi}\,\varphi(\theta',\theta)\mathsf{F}(\varepsilon(x,\theta')),\qquad
    \mbox{a.e. } (x,\theta)\in \R^2.
\end{equation}
\end{defi}
The local TBA form is often referred to as the {\em local-density approximation} (LDA). Clearly, the TBA form is recovered from the local TBA form by choosing $\beta(x,\theta) = \beta(\theta)$ independent of $x$ for $|x|<1/2$, and $\beta(x,\theta) = +\infty$ for $|x|\geq 1/2$.

\medskip
{\bf Extension: finite volumes.} It is not known in general if a TBA-like equation describes the thermodynamics of integrable systems in finite volumes. However, as we demonstrate later, it is possible to derive the finite-volume version of TBA for the semiclassical Bethe systems that we study here. We conjecture that the form of the finite-volume TBA equations that we obtain may also be valid in quantum integrable systems.

\begin{defi}\label{defiTBAfinite}
Consider the setup of Definition \ref{defiTBAlocal}. The finite-volume specific free energy, Eq.~\eqref{eq:freeenergyfinite}, is said to take the finite-volume TBA form if
\begin{equation}\label{fTBAL}
    f_L=\frac{1}{L}\int_{\mathbb{R}^2}\frac{\dd x\dd \theta}{2\pi}\, \mathsf{F}(\varepsilon_L(x,\theta))
\end{equation}
where the finite-volume pseudo-energy $\varepsilon(x,\theta)$ solves
\begin{equation}\label{eq:pseudoenergyfinitevolume}
    \varepsilon_L(x,\theta)=\beta(x/L,\theta)+\int_{\mathbb{R}}\frac{\dd x'\dd \theta'}{2\pi} \kappa_x(x-x',\theta,\theta')\mathsf{F}(\varepsilon_L(x',\theta')),\qquad
    \mbox{a.e. }(x,\theta)\in \R^2.
\end{equation}
Here $\kappa(x-x',\theta,\theta')$ is some differentiable function that may depend on $L$, and that reproduces the scattering shift $\varphi(\theta',\theta)$ at infinite volumes,
\beq\label{eq:limkappa}
	\lim_{x\to\infty}\lim_{L\to\infty}(\kappa(x,\theta,\theta')
	- \kappa(-x,\theta,\theta')) = \varphi(\theta',\theta).
\eeq
\end{defi}

We will show that for any semiclassical Bethe system, the finite-volume and infinite-volume specific free energies for chemical potential function $\beta(x,\theta)$ take the finite-volume and local TBA forms, with $\kappa(x,\theta,\theta') = \psi_{\theta}(x,\theta-\theta')$, under appropriate constraints and for a large family of functions $\beta(x,\theta)$.

\section{Proofs}\label{sectproofs}

In this section, we provide the proof of Theorems \ref{thsummary}, \ref{thsummaryttbar} and \ref{thsummaryTBA}.

\subsection{Proof of Theorem \ref{thsummary}a: trajectories}\label{ssectproofa}

We will show the following, which implies Theorem \ref{thsummary}a.
\begin{theorem}\label{thetrajectoriesfull}
The map $\omega:(\bs y,\bs\theta)\mapsto (\bs x,\bs p)$ defined by Eqs.~\eqref{cba1}, \eqref{cba2} is a $C^1$ diffeomorphism of $\R^{2N}$; further, fixing $\bs\theta$ the resulting map $\bs y\mapsto \bs x$ is a $C^1$ diffeomorphism of $\R^N$, and fixing $\bs x$ the resulting inverse map $\bs p\mapsto \bs\theta$ is also $C^1$ diffeomorphism of $\R^N$. Thus the trajectories $t\mapsto (\bs x(t),\bs p(t))$ exist and are unique, and their coordinates are continuously differentiable functions of $t$. Further, for every $(\bs y,\bs\theta)\in\R^{2N}_{\neq}$, there is a time $T>0$ such that for all $|t|\geq T$, trajectories are linear with  momenta $\bs p(t) = \bs \theta$.
\end{theorem}
\proof
Consider the generating function \eqref{generatingsummary},
\beq\label{generating}
	\Phi(\bs x,\bs\theta) =
	\sum_i x_i\theta_i
	+\frc12\sum_{i\neq j} \psi(x_i-x_j,\theta_i-\theta_j).
\eeq
The matrix (continuous in $\bs x,\bs\theta$)
\beq\label{Gamma}
	\Gamma_{ij}(\bs x,\bs \theta):=\p_{x_i}\p_{\theta_j}\Phi(\bs x,\bs \theta),
\eeq
is strictly positive, uniformly for $(\bs x,\bs 
\theta)\in \R^{2N}$. Indeed,
\beq\label{Gamma_2}
	\Gamma_{ij} = \delta_{ij}\Big(1
	+
	\sum_{k\neq i}\psi_{x\theta}(x_i-x_k,\theta_i-\theta_k)
	\Big)
	-
	(1-\delta_{ij})\psi_{x\theta}(x_i-x_j,\theta_i-\theta_j)
\eeq
wherefore for any $\bs v\in\R^N$ with $\bs v^2=1$ we have
\beqa
	\sum_{ij}
	\Gamma_{ij}v_iv_j &=&
	1 + \sum_{ij}
	v_i(v_i-v_j)
	\psi_{x\theta}(x_i-x_j,\theta_i-\theta_j)\n
	&=&
	1 + \frc12\sum_{ij}
	(v_i-v_j)^2
	\psi_{x\theta}(x_i-x_j,\theta_i-\theta_j)
	\ \geq\ 1.
 \label{eq:Gammavv}
\eeqa
We note that the matrix $\Gamma$ is related as the Gaudin matrix in quantum integrable systems \cite{gaudinbook}.

Consider the equations \eqref{cba1}, written in the form $y_i = \p_{\theta_i}\Phi(\bs x,\bs \theta)$ (see Eq.~\eqref{yeq}). By positivitity and continuity of $\Gamma$, they have a unique solution $\bs x$ for all $(\bs y,\bs \theta)\in\R^{2N}$, and the map $\bs y\mapsto \bs x$ (for every fixed $\bs\theta\in\R^N$) is a $C^1$ diffeomorphism of $\R^N$. Further, the map $(\bs y,\bs\theta)\mapsto \bs x$ defined by solving \eqref{cba1} is also in $C^1(\R^{2N})$, by the implicit function theorem. Directly computing $\bs p$ from \eqref{cba2} we also find that the map $(\bs x,\bs\theta)\to \bs p$ is in $C^1(\R^{2N})$. Thus we have constructed the map $\omega \in C^1(\R^{2N})$:
\beq
    \begin{aligned}
        \omega \quad: \quad \R^{2N}&\to\R^{2N} \\ (\bs y,\bs\theta)&\mapsto (\bs x,\bs p).
    \end{aligned}
\eeq
obtained by first computing $\bs x$ from $\bs y$ and $\bs\theta$ using \eqref{cba1}  (say at $t=0$) and then $\bs p$ from $\bs x$ and $\bs\theta$ using \eqref{cba2} (at $t=0$).

By repeating the same argument using \eqref{peq}, but replacing $\bs \theta \to \bs x$ and $\bs y \to \bs p$, we analogously find that \eqref{cba2} has a unique solution $\bs \theta$ for given $\bs x$ and $\bs p$, the resulting map $\bs p \mapsto \bs \theta$ is a $C^1$ diffeomorphism for every fixed $\bs x \in \mathbb{R}^N$, and $(\bs x,\bs p) \mapsto \bs \theta$ is continuously differntiable. This gives us $\omega^{-1}\in C^1(\R^{2N})$, the constructive inverses of $\omega$,
\beq
    \begin{aligned}
        \omega^{-1} \quad: \quad \R^{2N}&\to\R^{2N} \\ (\bs x,\bs p)&\mapsto (\bs y,\bs \theta),
    \end{aligned}
\eeq
by computing $\bs \theta$ from $\bs x$ and $\bs p$ using \eqref{cba2} and subsequently $\bs y$ via \eqref{cba1}. Hence, we conclude that $\omega$ is a $C^1$ diffeomorphism of $\R^{2N}$.


This shows the first sentence of Theorem \ref{thetrajectoriesfull}.

The second sentence follows easily: As $\bs y(t) = \bs y +\bs \theta t$, the resulting $x_i(t)$ are continuously differentiable in $t$. Further, from \eqref{cba2}, $p_i(t)$ (obviously unique) are also continuously differentiable in $t$.

Finally, by the conditions \eqref{condpsioverview}, the function $\psi_\theta(x,\theta)$ is constant in $x$ for all $|x|\geq \ep$,
\beq\label{psithetaaymp}
    \psi_\theta(x,\theta) = \psi_\theta^{\sgn(x)}(\theta)\qquad(|x|\geq \ep).
\eeq 
As it is also continuous in $x$ (and in fact strictly non-decreasing by \eqref{condpsioverview}), it is bounded,
\beq
    m(\theta) := \max\{|\psi_\theta(x,\theta)|:x\in\R\} <\infty.
\eeq
Therefore, from \eqref{cba1} replacing $\bs y$ by its time-evolved form $\bs y + \bs \theta t$ (and keeping implicit only the particles' coordinates' argument $t$),
\beq
    |y_i-x_i+\theta_i t| \leq \sum_{j\neq i}m(\theta_i-\theta_j) =:M_i.
\eeq
Hence, for all $\bs \theta$ with $\theta_i\neq \theta_j\,\forall\,i\neq j$, and for all $|t|\geq T$ with
\beq
    T = \max\Big\{\frc{|y_i-y_j|+ M_i+M_j + \ep}{|\theta_i-\theta_j|}: i,j\in\I_N,\,i\neq j\Big\},
\eeq
we have for every $i\neq j$
\beq
    |x_i-x_j| \geq |y_i-y_j+(\theta_i-\theta_j)t| - M_i-M_j \geq |\theta_i-\theta_j|\, |t|
    - |y_i-y_j| - M_i-M_j
    \geq \ep.
\eeq
For all $t\geq T$, clearly $x_i-x_j$ does not cross zero, hence its sign is determined by the limit $t\to\infty$; similarly for $t\leq-T$. Thus for such times, with \eqref{cba1} at time $t$ we find
\beq
    \sgn(x_i-x_j) = \sgn(\theta_i-\theta_j)\sgn(t)\qquad (|t|\geq T).
\eeq
Hence in \eqref{cba1} and \eqref{cba2} we may replace $\psi_\theta(x_i-x_j,\theta_i-\theta_j)$ by $\psi_\theta^{\sgn(\theta_i-\theta_j)\sgn(t)}(\theta_i-\theta_j)$ and $\psi_x(x_i-x_j,\theta_i-\theta_j)$ by 0. This shows the last sentence of Theorem \ref{thetrajectoriesfull}. Specifically, for every $(\bs y,\bs\theta)\in\R^{2N}_{\neq}$, we have
\begin{align}\label{xipiasymp}
    \forall\ |t|>T\ :\quad x_i &= y_i + \theta_i t - \sum_{j\neq i}
    \psi_\theta^{\sgn(\theta_i-\theta_j)\sgn(t)}(\theta_i-\theta_j)\\
    p_i &= \theta_i.
\end{align}
This completes the proof. \eproof

\subsection{Proof of Theorem \ref{thsummary}b: Liouville integrability and finite-range interaction}\label{ssectproofb}

We show that the Hamiltonian $H_\psi$ from Eq.~\eqref{Hpsi} is Liouville integrable, with $N$ functionally independent conserved quantities $Q_a\in C^1(\R^{2N})$ in involution, that can be taken as
\beq
    Q_a(\bs x,\bs p) = \sum_i \theta_i(\bs x,\bs p)^a
\eeq
for $a=1,2,\ldots,N$.

These are indeed functionally independent, as for all $\bs x$ such that $|x_i-x_j|\geq \ep\,\forall\,i\neq j$, we have $\theta_i = p_i$, and within this region of $\R^{2N}$, the functions $Q_a$ are clearly functionally independent. By \eqref{yeq} and \eqref{peq}, we see that the transformation of coordinate $\omega^{-1}$ is a canonical transformation, generated by the mixed generating function $\Phi\in C^2(\R^{2N})$. Therefore
\beq\label{ythetacanonical}
    \{y_i,\theta_j\} = \delta_{ij},\quad
    \{y_i,y_j\} = \{\theta_i,\theta_j\}=0.
\eeq
Hence $Q_a$'s are in involution. Further, from \eqref{ythetacanonical} the Hamiltonian \eqref{Hpsi} gives
\beq
    \{y_i,H_\psi\} = \theta_i,\quad
    \{\theta_i,H_\psi\} = 0.
\eeq
Taking $H_\psi$ as the generator of the flow in time, the solutions are $y_i(t) = y_i + \theta_i t$ and $\theta_i(t) = 0$. Hence the trajectories defined by \eqref{cba1}, \eqref{cba2} indeed are generated by $H_\psi$. Thus the first sentence of Theorem \ref{thsummary}b holds.

Consider the quasipotentials
\beq
    V_a(\bs x,\bs p) = \sum_i (\theta_i(\bs x,\bs p)^a - p_i^a),\quad V_a \in C^1(\R^{2N}).
\eeq
According to Definition \ref{defilocal}, in order to show the second sentence of Theorem \ref{thsummary}b, we only need to show that $V_a$ has finite range uniformly, that is all $V_a$'s satisfy \eqref{condpotential} where $V_a^{(n)}$ do not depend on $N$. Then, with $E_a = p^a$, we find that $H$ is Liouville integrable with finite range interaction $V_\psi = V_2/2$.

\begin{lemma}\label{lemtheta}
Suppose that for a bipartition $I,\,J$ of $\{1,\ldots,N\}$ (that is $I\cap J = \emptyset,\,I\cup J = \{1,\ldots,N\}$), we have $|x_i-x_j|>\ep$ for all $i\in I,\,j\in J$.  Then $\p_{x_i}\theta_j(\bs x,\bs p)=\p_{p_i}\theta_j(\bs x,\bs p)=0$ for all $i\in I,\,j\in J$; if $J = \{j\}$ has a single element, then $\theta_j(\bs x,\bs p) = p_j$; and in general $\theta_i = \theta_i(\bs x_I,\bs p_I)$ and $\theta_j = \theta_j(\bs x_J,\bs p_J)$ ($i\in I,\,j\in J$) take the form of the asymptotic momenta of the semiclassical Bethe system restricted to $I$ and $J$, respectively:
\beq\begin{aligned}\label{cba2separated}
    p_i &= \theta_i + \sum_{i'\in I,i'\neq i}
    \psi_x(x_i-x_{i'},\theta_i-\theta_{i'})\qquad
    \forall i\in I,\\
    p_j &= \theta_j + \sum_{j'\in J,j'\neq j}
    \psi_x(x_j-x_{j'},\theta_j-\theta_{j'})\qquad
    \forall j\in J,\\
    \end{aligned}
\eeq
\end{lemma}
\proof
From the last equation of \eqref{condpsioverview}, for $i\in I$ the right-hand side of Eq.~\eqref{cba2} becomes the right-hand side of the first of Eqs.~\eqref{cba2separated}; likewise for the second of Eqs.~\eqref{cba2separated}. The general argument in the proof of Theorem \ref{thetrajectoriesfull} holds individually for each of these equations, which thus have unique $C^1(\R^{2|I|})$ and $C^1(\R^{2|J|})$ solutions $\theta_i(\bs x_I,\bs p_I)$ and $\theta_j(\bs x_J,\bs p_J)$, respectively.
\eproof

This shows the third sentence of Theorem \ref{thsummary}b.

\begin{corol}
For each $a=1,2,\ldots$, the function $V_a(\bs x,\bs p)$ satisfies the finite-range condition \eqref{condpotential} uniformly in $N$.
\end{corol}
\proof Condition \eqref{condpotential} for each $V_a(\bs x,\bs p)$ follows from Lemma \ref{lemtheta} with
\beq
	V_a^{(I)}(\bs x_I,\bs p_I) = \sum_{i\in I} (\theta_i(\bs x,\bs p)-p_i)^a,
\eeq
which clearly does not depend on $N$.
\eproof

This completes the proof of the second sentence of Theorem \ref{thsummary}b.

\subsection{Proof of Theorem \ref{thsummary}c: elastic, factorised scattering}

We consider Definition \ref{defiscattering}. We will show the property of elastic, factorised scattering, with the dense subspace of initial condition $\mathcal S' = \omega(\R^{2N}_{\neq})$. This is clearly invariant under time evolution, as the condition $\theta_i\neq \theta_j\,(i\neq j)$ is preserved since $\theta_i$'s are constant. Relations \eqref{xipiasymp} immediately implies part (1) of Definition \ref{defiscattering}. In particular \eqref{asymptotic} holds with
\beq\label{xpm}
    \sigma^\pm = \1,\quad
    x_i^\pm = y_i - \sum_{j\neq i}
    \psi_\theta^{\pm\sgn(\theta_i-\theta_j)}(\theta_i-\theta_j),
\eeq
and thus $H_\psi$ is a tracer dynamics.

From this we evaluate
\beq
    x_i^+ - x_i^- = -\sum_{j\neq i}
    \psi_\theta^{\sgn(\theta_i-\theta_j)}
    (\theta_i-\theta_j) +
    \psi_\theta^{-\sgn(\theta_i-\theta_j)}
    (\theta_i-\theta_j)
    = - \sum_{j\neq i}\sgn(\theta_i-\theta_j)\varphi(\theta_i-\theta_j)
\eeq
which implies part (2) of Definition \ref{defiscattering} with $\varphi(\theta,\theta') = \varphi(\theta-\theta')$.

As $\omega$ is a bijection, the map $\omega^{-1}:(\bs x(0),\bs p(0))\mapsto (\bs y,\bs\theta)$ is a bijection from $\mathcal S'$ to $\R^{2N}_{\neq}$. From \eqref{xpm}, for every fixed $\bs\theta:\,\theta_i\neq \theta_j\,(i\neq j)$, we have a bijection $\R^N\ni \bs y\mapsto\bs x^-\in\R^N$. Thus there is a bijection $\R^{2N}_{\neq}\ni(\bs y,\bs\theta)\mapsto(\bs x^-,\bs\theta)\in \R^{2N}_{\neq}$, and hence the map $(\bs x(0),\bs p(0))\mapsto (\bs x^-,\bs\theta)$ is a bijection, which shows part (3) of Definition \ref{defiscattering}.

Hence we have shown that $H_\psi$ satisfies the property of elastic and factorised scattering, and is a tracer dynamics, with two-body scattering shift $\varphi(\theta,\theta') = \varphi(\theta-\theta')$.

It remains to be shown that $H_\psi$ is a finite-range tracer dynamics, Definition \ref{defifinitetracer}. This follows immediately from Lemma \ref{lemtheta}.

\subsection{Proof of Theorem \ref{thsummaryttbar}: generalised $T\b T$-deformation}

Let $\psi\in C^2(\R^2)$ satisfy \eqref{condpsioverview}, and consider the Hamiltonian $H_\psi$. We will show that $\psi^\lambda := \lambda \psi \in C^2(\R^2)$, which satisfies $\psi^1=\psi$, $\psi^0 = 0$, and conditions \eqref{condpsioverview} for all $\lambda\in\mathbb R$, is such that $H_\lambda(\bs x,\bs p):= H_{\psi^\lambda}(\bs x,\bs p)$ is differentiable in $\lambda$ and satisfies \eqref{Hhat} for $\mathcal S_\lambda' = \omega^\lambda(\R^{2N}_{\neq})$ and
\beq
    w(x,\theta) = \frc12 \psi_x(x,\theta).
\eeq
By Theorems \ref{thsummary}b and \ref{thsummary}c, each $H_\lambda$ is Liouville integrable with finite-range interaction, satisfies the elastic, factorised scattering property with $\mathcal S_\lambda' = \omega^\lambda(\R^{2N}_{\neq})$, and is a tracer dynamics with finite-range interaction. Thus, according to Definition \ref{defigenTTbar}, $H_\psi$ is a generalised $T\b T$-deformation of $H_{\psi^0} = H_{\rm free}$.

We note that the deformation function $w\in C^1(\R^2)$ satisfies the properties
\beq\label{w_def}
	w_\theta(x,\theta)\geq 0,\quad
	w(-x,-\theta) = -w(x,\theta),\quad
	w(x,\theta) = 0 \ \mbox{whenever $|x|>\ep$}
\eeq
and that one can write
\beq\label{psilambda}
    \psi^\lambda(x,\theta) = \lambda \int_{-\infty}^\infty \dd x'\,\sgn(x-x')w(x',\theta).
\eeq
In particular, the two-body scattering shift for $H_\psi$ is related to the deformation function as
\beq
    \varphi(\theta) = 2\int \dd x\,w_\theta(x,\theta).
\eeq
Along the flow, the scattering shift is simply re-scaled by $\lambda$.

We evaluate the asymptotic momenta $\theta_i^\lambda(\bs x,\bs p)$ from \eqref{cba2} (here and below, we use the upper index $\lambda$ to represent phase space functions associated to the Hamiltonian $H_\lambda = H_{\psi^\lambda}$),
\beq\label{cba2ttbar}
    p_i = \theta_i^\lambda + \lambda\sum_{j\neq i}
    \psi_x(x_i-x_j,\theta_i^\lambda-\theta_j^\lambda),
\eeq
which by Theorem \ref{thsummary}a has a solution for all $\lambda\in\R$. Differentiability of $H_\lambda(\bs x,\bs\theta)$ in $\lambda$ follows from \eqref{cba2ttbar} and the implicit function theorem.

It is well known that $T\b T$-deformations are generated by a certain bilinear in the conserved densities \cite{Pozsgay2020,10.21468/SciPostPhys.9.5.078} (see \cite{cardy2022t} for the case of classical particles). Adapting this to our case, we consider the observable
\beqa
	X^\lambda(\bs x,\bs p) &=& -\int \dd x\dd y \dd z\dd \theta\dd\alpha\,
	\Theta(z-x) w(z-y,\theta-\alpha)q_{\theta}^\lambda(x)q_{\alpha}^\lambda(y)\n
	&=& -\sum_{ij} \int_{x_i}^\infty \dd z\,w(z-x_j,\theta_i^\lambda(\bs x,\bs p)-\theta_j^\lambda(\bs x,\bs p))\label{X}
\eeqa
where $\Theta(z)$ is the Heaviside theta-function. We now show that
\beq
	\{H_\lambda,X^\lambda\} = \Delta H_\lambda
\eeq
where $\Delta H_\lambda$ is given by the right-hand side of \eqref{Hhat}. This is a relatively standard calculation; the bilinear form of $X^\lambda$, the first line in \eqref{X}, gives the correct $T\b T$-deformation thanks to the conservation laws. Here, in fact, we may directly use the second line in \eqref{X}, along with $\{x_j,H_\lambda\} = \dot x_j^\lambda$ and $\{\theta_i^\lambda,H_\lambda\}=0$:
\beq
	\{H_\lambda,X^\lambda\} = -\sum_{ij} \dot x_j^\lambda \int_{x_i}^\infty \dd x\,w_x(x-x_j,\theta_i^\lambda-\theta_j^\lambda) - \sum_{ij} \dot x_i^\lambda w(x_i-x_j,\theta_i^\lambda-\theta_j^\lambda)
	= \Delta H_\lambda.
\eeq
Then, in order to prove the theorem, we must simply show that
\beq
	\{H_{\lambda},X^\lambda\} = \frc{\p}{\p\lambda}
	H_{{\lambda}}.
\eeq

We may use the form of the Hamiltonian in terms of asymptotic momenta
\beq
	H_{\lambda}(\bs x,\bs p) = \frc12\sum_i \big(\theta^\lambda_i(\bs x,\bs p)\big)^2
\eeq
Thus by the Leibniz rule it is sufficient to show
\beq\label{theta_identity}
	\{\theta^\lambda_i,X^\lambda\} = \frc{\p}{\p\lambda}\theta^\lambda_i.
\eeq
We denote $f_{ij}:=\{\theta_i^\lambda,x_j\}$. The left-hand side of \eqref{theta_identity} can be written as
\beq
\{\theta_i^\lambda,X^\lambda\}
	=\sum_{jk} (\{\theta_i^\lambda,x_j\} - \{\theta_i^\lambda,x_k\})
	w(x_{jk},\theta_{jk}^\lambda)=2\sum_{jk} f_{ij}
	w(x_{jk},\theta_{jk}^\lambda),
\eeq
where we use the short-hand notation $x_{jk} = x_j-x_k$, etc., and the last equality follows from $w(-x,-\theta)=-w(x,\theta)$ as per Eq.~\eqref{w_def}.
Using Eq.~\eqref{psilambda} and Eq.~\eqref{cba1}, we have
\begin{align}
    -\delta_{ij}=f_{ij}+2\lambda\sum_{k\neq j}(f_{ij}-f_{ik})w_\theta(x_{jk},\theta^\lambda_{jk}) 
    =\sum_kf_{ik}(W_{kj}-V_{kj}),
\end{align}
which gives $f_{ij}=-(W-V)^{-1}_{ij}$ where
\begin{equation}
    W_{ij}\equiv\delta_{ij}\left(1+\sum_{k} V_{ik}\right),\quad V_{jk}=V_{kj}:=(1-\delta_{jk})2\lambda w_\theta(x_{jk},\theta^\lambda_{jk}).
\end{equation}
The matrix $W-V$ is strictly positive definite by a calculation similar to that done for $\Gamma$ in Subsection \ref{ssectproofa}, hence is invertible. We therefore obtain
\begin{equation}
    \{\theta^\lambda_i,X^\lambda\} =-2\sum_{jk}(W-V)^{-1}_{ij}w(x_{jk},\theta_{jk}^\lambda).
\end{equation}
Next we compute $\p\theta^\lambda_i/\p\lambda$ using Eq.~\eqref{psilambda} and Eq.~\eqref{cba2}, yielding
\begin{equation}
    0=\frac{\p\theta^\lambda_i}{\p\lambda}+2\sum_{k\neq i}w(x_{ik},\theta_{ik}^\lambda)+2\lambda\sum_{k\neq i}\frac{\p\theta^\lambda_{ik}}{\p\lambda}w_\theta(x_{ik},\theta^\lambda_{ik}),
\end{equation}
This can be easily solved as
\begin{equation}
    \frac{\p\theta^\lambda_i}{\p\lambda}=-2\sum_{jk}(W-V)^{-1}_{ij}w(x_{jk},\theta_{jk}^\lambda),
\end{equation}
thereby Eq.~\eqref{theta_identity} is established.

\subsection{Proof of Theorem \ref{thsummaryTBA}: thermodynamic Bethe ansatz}

We now compute the specific free energy of the semiclassical Bethe systems. Recall that by Theorem \ref{thsummary}c, these are finite-range tracer dynamics. We consider the partition function \eqref{eq:ggetracerlda}, with space-varying chemical potentials; as discussed, this is the most general formulation. Thus we consider 
\begin{equation}
\label{ggemostgeneral}
    Z_L=\sum_{N=0}^\infty\frac{1}{(2\pi)^NN!}\int\prod_{j=1}^N\dd x_j\dd p_j\,e^{-\int \dd x \dd\theta\, \beta(x/L,\theta)q_\theta(\bs x,\bs p;x)}
\end{equation}
for a semiclassical Bethe system, for some finite $L>0$ and some Lebesgue measurable function $e^{-\beta} : \R^2\to [0,\infty)$ (note that $\beta(x,\theta)\in \R\cup\{+\infty\}$). We are looking to evaluate
\begin{equation}\label{eq:fLproof}
    f_L = - \frc1L\log Z_L\quad\mbox{and}\quad
    f = \lim_{L\to\infty}f_L.
\end{equation}

Theorem \ref{thsummaryTBA} is a corollary of the Theorems \ref{theoTBA} and \ref{theoTBA2}.

\begin{theorem}\label{theoTBA}
Consider a semiclassical Bethe system. Assume the following:
\begin{equation}
\label{eq:integralphi}
	\int \dd\theta\, \varphi(\theta) = \varphi_{\rm tot} < \infty,
\end{equation}
\begin{equation}\label{eq:boundbeta}
    \int \dd x\dd \theta\,
    e^{-\beta(x,\theta)}
    = F < \infty,
\end{equation}
\begin{equation}\label{eq:boundbetasup}
	\sup_{(x,\theta)\in\R^2} e^{-\beta(x,\theta)} =
	e^{-\beta_{-}} <\infty
\end{equation}
(each of these quantities is non-negative) and
\begin{equation}\label{eq:boundCE_2}
    A:=\frc{e^{1-\beta_{-}}\varphi_{\rm tot}}{2\pi} < 1.
\end{equation}
For every $L>0$, we have $Z_L<\infty$ and $f_L\leq 0$; and $\sup_{L>0}|f_L|<\infty$. In particular,
\beq
	\sup_{L>0}|f_L| \leq  \frc{F}{2\pi}\frc{A}{1-A}.
\eeq
Further, $f_L$ takes the finite-volume TBA form for Maxwell-Boltzmann statistics $\mathsf F(\varep) = -e^{-\varep}$ and two-body scattering shift $\varphi(\theta-\theta')$ (with $\kappa(x,\theta,\theta') = \psi_{\theta}(x,\theta-\theta')$; Definition \ref{defiTBAfinite}).
\end{theorem}

\proof
Note that the Jacobian produced as a result of switching from $(\bs x,
\bs p)$ coordinates to $(\bs x,
\bs \theta)$ is nothing other than the Gaudin determinant $|\Gamma(\bs x,
\bs \theta)|$. We thus rewrite the partition function Eq.~\eqref{ggemostgeneral} as
\begin{equation}\label{eq:gge2}
    Z_L=\sum_{N=0}^\infty\frac{1}{(2\pi)^NN!}\int\prod_{j=1}^N\dd x_j\dd \theta_j\,|\Gamma(\bs x,
\bs \theta)|e^{-\sum_{i=1}^N \beta(x_i/L,\theta_i)}.
\end{equation}
It is clear that sharp boundaries (giving a system supported on the finite volume $[-L/2,L/2]$) can be obtained by setting $\beta(x,\theta) = \infty$ for all $|x|
\geq 1/2$; this still satisfies conditions \eqref{eq:boundbeta}, \eqref{eq:boundbetasup}, \eqref{eq:boundCE_2}.

The integrand, and every term in the series, are non-negative. Hence in order to show absolute convergence, we only need to show convergence for some re-organisation of the terms. We will use a combinatoric argument to re-organise the terms appropriately, from which we will both show convergence and obtain the TBA form.

At first glance the integral does not seem particularly simple, but it turns out that a graph theoretic interpretation of the Gaudin determinant allows us to evaluate it exactly. The same observation was in fact made already for quantum integrable systems in \cite{10.1007/978-981-13-2179-5_6}. Since the calculation is almost identical, we omit the detail of the computation, and outline the steps in Appendix \ref{app:matrixtree}.

The calculation is based on expressing the Gaudin matrix as $\Gamma_{ij}=
\delta_{ij}+L_{ij}$, where $L_{ij}$ is a Laplacian matrix (a matrix with zero row sums)
\beq\label{eq:laplace}
	L_{ij} = \delta_{ij}
	\sum_{k\neq i}\psi_{x\theta}(x_{ik},\theta_{ik})
	-
	(1-\delta_{ij})\psi_{x\theta}(x_{ij},\theta_{ij}),
\eeq
where we use again $x_{ij}=x_i-x_j$ and $\theta_{ij}=\theta_i-\theta_j$. We then have
\begin{equation}
    |\Gamma|=\sum_{\alpha\subset\{1,\cdots,N\}}\mathcal{L}(\alpha|\alpha),
\end{equation}
where $\mathcal{L}(\alpha|\alpha)$ is the principal minor obtained by removing the rows and column $\alpha$ from $L_{ij}$ (and $\mathcal{L}(\alpha|\alpha) = 1$ for $\alpha = \{1,\ldots,N\}$). The crucial insight of \cite{10.1007/978-981-13-2179-5_6} is to notice that the matrix tree theorem allows us to evaluate the principal minors by appropriate sums over graphs, providing a combinatorial interpretation of $|\Gamma|$.

The result (see Appendix \ref{app:matrixtree}) is an expression for the term $N$ in the series \eqref{eq:gge2}, as a sum over all distinct topological rooted forests with $N$ vertices. A topological  rooted forest with $N$ vertices is the (unordered) juxtaposition of topological rooted trees $m$ with $N_m$ vertices and $\sum_m N_m = N$; and a topological rooted tree with $N'$ vertices is the equivalence class, under permutations of non-root vertices $\{2,3,\ldots,N'\}$ amongst themselves, of a rooted spanning tree for the complete graph with $N'$ vertices labelled $1,2,\ldots,N'$, with root at vertex 1. In their evaluations, the forest and trees come with appropriate symmetry factors. Organising the (countable) set of all topological rooted trees with a number $m=1,2,\ldots$ (say with non-decreasing number of vertices), and denoting $N_m\geq 1$ the number of vertices for tree $m$, the result is
\begin{equation}\label{termforest}
    \frc1{(2\pi)^NN!}
    \int\prod_{j=1}^N\dd x_j\dd \theta_j\,|\Gamma(\bs x,
\bs \theta)|e^{-\sum_{i=1}^N \beta(x_i/L,\theta_i)}
    =
    \sum_{n_1,n_2,\ldots=0\atop
    \sum_m n_m N_m = N}^\infty
    \prod_m \frc1{n_m!}V_m^{n_m}.
\end{equation}
Here, each term is the value of a topological rooted forest with $\sum_m n_m$ roots (trees); note that for finite $N$ there is a finite number of terms. The number $n_m$ is the multiplicity of topological rooted tree $m$ in the forest, and the tree $m$ has value
\begin{equation}
    V_m=\frac{1}{S_m}\frc1{(2\pi)^{N_m}}\int\prod_{i=1}^{N_m}\dd x_i\dd \theta_i\,e^{-\sum_{i=1}^{N_m}\beta(x_i/L,\theta_i)}\prod_{j=1}^{N_m-1}\psi_{x\theta}(x_{{\rm e}_j},\theta_{{\rm e}_j}),
\end{equation}
with ${\rm e}_j = \{i_{j,1},i_{j,2}\},\,j=1,\ldots,N_m-1$ the set of edges (sets of two vertex labels) of one representative of the equivalence class, and $S_m$ the tree's symmetry factor. The last product in the above equation is 1 if $N_m=1$. In fact, because of the full symmetry under permutation of integration variables, it is not necessary at this stage to consider rooted trees (that is, we could sum over unrooted forests), but it will be convenient below in order to obtain the TBA form.

We note that the total number of spanning trees for the complete graph of $N$ vertices is $N^{N-2}$ according to Cayley's formula \cite{cayley_2009}, and that the total number of trees in the equivalence class $m$ is $(N_m-1)!/S_m$ (this formula in fact defines the symmetry factor $S_m$; we remark that there are $(N_m-1)!$ permutations of non-root vertices). Thus
\begin{equation}\label{eq:countingtrees}
    \sum_{m:N_m=N} \frc1{S_m}
    = \frc{N^{N-1}}{N!}.
\end{equation}

Expression \eqref{termforest} allows us to re-organise the series \eqref{eq:gge2} as
\begin{equation}\label{eq:ZlimiN}
    Z_L = \lim_{N\to\infty} \sum_{k=0}^\infty \frc1{k!} \Bigg(\sum_{m: N_m\leq N} V_m\Bigg)^k
    = \lim_{N\to\infty}
    \exp \Bigg[\sum_{m: N_m\leq N} V_m\Bigg].
\end{equation}
We see that the specific free energy $f = L^{-1} \log Z$ is indeed non-positive.

From the above, we may now analyse convergence of the partition function. It is convenient to bound the result of summing over all topological rooted trees with ``fixed root", that is writing
\begin{equation}\label{Vmintegral}
    V_m = \frc1{2\pi} \int \dd x_1 \dd \theta_1\,
    e^{-\beta(x_1/L,\theta_1)}
    V_m(x_1,\theta_1)
\end{equation}
where
\begin{equation}\label{eq:Vmx1t1}
    V_m(x_1,\theta_1) =
    \frac{1}{S_m}\frc1{(2\pi)^{N_m-1}}\int\prod_{i=2}^{N_m}\dd x_i\dd \theta_i\,e^{-\sum_{i=2}^{N_m}\beta(x_i/L,\theta_i)}\prod_{j=1}^{N_m-1}\psi_{x\theta}(x_{{\rm e}_j},\theta_{{\rm e}_j}).
\end{equation}
We sequentially bound the integral over $x_{j},\,\theta_{j}$ for $j=N_m$, $j=N_m-1$, up to $j=2$, by using non-negativity of $\psi_{x\theta}(x,\theta)$ from \eqref{condpsioverview}, the definition \eqref{defphioverview},  finiteness of the total scattering shift \eqref{eq:integralphi}, and the boundedness condition \eqref{eq:boundbetasup} on $e^{-\beta(x,\theta)}$:
\begin{equation}
    0\leq \int\dd x
    \dd\theta\,
    e^{-\beta(x/L,\theta)}
    \psi_{x\theta}(x'-x,\theta'-\theta)\leq
    e^{-\beta_{-}}\varphi_{\rm tot} = 2\pi C
\end{equation}
where $C=Ae^{-1}$. Therefore
\beq
    V_m(x,\theta) \leq \frc{C^{N_m-1}}{S_m}
\eeq
for all $x,\theta$. Eq.~\eqref{eq:countingtrees} then gives
\begin{equation}\label{eq:boundforVm}
    \sum_{m:N_m=N} V_m(x,\theta)
    \leq
    C^{N-1}
    \frc{N^{N-1}}{N!} =: a_N.
\end{equation}
Note that $\frac{a_{N+1}}{a_N} = C\left(1+\frac{1}{N}\right)^{N-1} \leq C\left(1+\frac{1}{N}\right)^N \leq Ce=A$ and therefore $a_N \leq A^N$. Since $A<1$ by \eqref{eq:boundCE_2} we thus find that
\begin{equation}
    Y(x,\theta) := e^{-\beta(x/L,\theta)}\sum_m V_m(x,\theta)
\end{equation}
converges, with
\begin{equation}
\label{eq:boundY}
    0<Y(x,\theta) = e^{-\beta(x/L,\theta)}\sum_{N=1}^\infty\sum_{m:N_m=N} V_m(x,\theta) \leq e^{-\beta(x/L,\theta)}
    \sum_{N=1}^{\infty}
    A^N
    = e^{-\beta(x/L,\theta)}
    \frc{A}{1-A}.
\end{equation}
Therefore, we may evaluate the limit in \eqref{eq:ZlimiN} from \eqref{Vmintegral} by evaluating the limit inside the integral: as the summands are non-negative, the pointwise limit $N\to\infty$ of the sum is a uniform upper bound and, by \eqref{eq:boundbeta}, it has finite integral, hence we may use the bounded convergence theorem. We find that
\begin{equation}\label{eq:resultZLproof}
    0\leq \log Z_L = \frc1{2\pi}\int\dd x \dd \theta\, Y(x,\theta) \leq \frc1{2\pi}
    \frc{LFA}{1-A}
\end{equation}
hence the limit \eqref{eq:ZlimiN} exists and we have absolute convergence of the partition function. On the right-hand side of \eqref{eq:resultZLproof}, the full $L$ dependence is explicit. Hence we find that $\log Z_L$ grows at most like $L$, thus the finite-volume specific free energy is bounded on $L>0$:
\beq\label{eq:Cboundspecial}
	\sup_{L>0}|f_L|
	\leq \frc1{2\pi}\frc{FA}{1-A}.
\eeq
This proves the bound expressed in the theorem.

One can reconstruct all topological rooted trees with a given root, by adjoining to this root $n=1$ edge, then $n=2$ edges, then $n=3$ edges, etc, and connecting to these edges the roots of all possible topological rooted trees, dividing by $n!$ for the symmetry. Thus we get a Schwinger-Dyson equation \cite{10.1007/978-981-13-2179-5_6},
\begin{align}
    Y(x,\theta)&=e^{-\beta(x/L,\theta)}\sum_{n=0}^\infty\frac{1}{2\pi n!}\left(\int\dd x'\dd \theta'\,\psi_{x\theta}(x-x',\theta-\theta')Y(x',\theta')\right)^n \n
    &=e^{-\beta(x/L,\theta)}\exp\left[\frc1{2\pi}\int\dd x'\dd \theta'\,\psi_{x\theta}(x-x',\theta-\theta')Y(x',\theta')\right].
    \label{Yequation}
\end{align}
Introducing the pseudo-energy $\varepsilon_L(x,\theta)$ defined by $Y(x,\theta)=: e^{-\varepsilon_L(x,\theta)}$, the last equation can be recast into the equation
\begin{equation}\label{eq:tba1}
    \varepsilon_L(x,\theta)=\beta(x/L,\theta)-\frc1{2\pi}\int\dd x'\dd \theta'\,\psi_{x\theta}(x-x',\theta-\theta')e^{-\varepsilon_L(x',\theta')}.
\end{equation}
Note that in the case of sharp boundaries, we have $\varep_L(x,\theta) = \infty$ for all $|x|\geq L/2$, thus the $x'$ integrand is supported $(-L/2,L/2)$.
Thus we have shown that the partition function takes the finite-volume local  TBA form both with and without sharp boundaries, for Maxwell-Boltzmann statistics and two-body scattering shift $\varphi(\theta-\theta')$, with $\kappa(x,\theta,\theta') = \psi_\theta(x,\theta-\theta')$, where Eq.~\eqref{eq:limkappa} holds thanks to \eqref{defphioverview} and Theorem \ref{thsummary}c. This completes the proof of Theorem \ref{theoTBA}. \eproof

\begin{theorem}\label{theoTBA2} In the context of Theorem \ref{theoTBA}, assume additionally the following. First, $e^{-\beta(x,\theta)}$ is continuous in $x$, uniformly for a.e.~$(x,\theta)\in\R^2$; that is, there exists a set $S\subset \R^{2}$ of Lebesgue measure zero, and $R:\R\to\R^+$ with $\lim_{u\to0} R(u) = 0$, such that
\beq\label{eq:uniformcontinuity}
	\Big|
	e^{-\beta(x+u,\theta)} - e^{-\beta(x,\theta)}
	\Big| \leq R(u)\quad \forall\,(x,\theta)\in\R^2\setminus S,\;u\in\R.
\eeq
Second,
\beq\label{eq:boundD}
	A<\frc{e}2\big(\sqrt{1+4e^{-1}}-1\big) = 0.77757288...
\eeq
(The bound \eqref{eq:boundD} implies \eqref{eq:boundCE_2}.) Then the specific free energy $f$ exists, and takes the local TBA form for Maxwell-Boltzmann statistics $\mathsf F(\varep) = -e^{-\varep}$ and two-body scattering shift $\varphi(\theta-\theta')$ (Definition \ref{defiTBAlocal}). Further, the solution $\varep(x,\theta)$ to the integral equation \eqref{pseudoenergydeflocal} (with $r=\infty$) exists for a.e.~$x\in\R$ and all $\theta\in\R$, is unique, and satisfies
\beq\label{eq:boundsolution}
	e^{-\varep(x,\theta)} < \frc{2\pi}{\varphi_{\rm tot}}\qquad
	\forall\;x,\theta\in\R.
\eeq
\end{theorem}

\proof
As \eqref{eq:boundD} implies $A<1$, \eqref{eq:boundCE_2} indeed holds. Recalling \eqref{eq:Cboundspecial}, we define
\beq\label{eq:beta0}
	\beta_{+} = \log\Big(\frc{A}{1-A}\Big)\in\R,
\eeq
and it will be convenient to define the following quantity, which evaluates as follows from the definition of $A$ in \eqref{eq:boundCE_2} and is bounded as follows by \eqref{eq:boundD},
\begin{equation}
    \label{eq:defDproof}
	D:=\frc{A^2e^{-1}}{1-A} = \frc{e^{\beta_{+}-\beta_{-}}\varphi_{\rm tot}}{2\pi}<1.
\end{equation}

Consider the scaled pseudo-energy  $\b\varep_L(x,\theta) = \varep_L(Lx,\theta) = - \log Y(Lx,\theta)$. For lightness of notation, we keep the dependence on $L$ of $\bar\varepsilon_L(x,\theta)$ implicit, writing instead $\b\varep(x,\theta)$. By \eqref{eq:boundY} and strict positivity of $Y(x,\theta)$ this is real and bounded as
\beq\label{eq:boundvarepsilon}
	\beta(x,\theta) - \beta_{+}  \leq \b\varep(x,\theta) < \infty.
\eeq
Also that by the bound \eqref{eq:boundbetasup}, $e^{-\varep(x,\theta)}$ is bounded from above:
\beq\label{eq:bounduniformvarep}
	\sup_{(x,\theta)\in\R^2,L>0} e^{-\b\varep(x,\theta)} \leq e^{\beta_{+}-\beta_{-}}
	<\infty.
\eeq
Defining $\Delta\b\varep(x,\theta) = \b\varep(x,\theta) - \beta(x,\theta)$, we also have the bound
\beq\label{eq:bounduniformdeltavarep}
	\sup_{(x,\theta)\in\R^2,L>0} e^{-\Delta\b\varep(x,\theta)} \leq e^{\beta_{+}}
	<\infty.
\eeq

By Theorem \ref{theoTBA} (see \eqref{eq:tba1} and \eqref{eq:resultZLproof}),
\begin{equation}\label{eq:tbabar}
    \b\varepsilon(x,\theta)=\beta(x,\theta)-\frc1{2\pi}\int\dd x'\dd \theta'\,L\psi_{x\theta}(Lx-Lx',\theta-\theta')e^{-\b\varepsilon(x',\theta')}
\end{equation}
and
\beq\label{eq:fLproof_2}
	f_L = -\int \frc{\dd x\dd\theta}{2\pi} e^{-\b\varep(x,\theta)}.
\eeq
We are looking to evaluate the limit as $L\to\infty$.

We show that $\Delta\b\varep(x,\theta)$ is continuous in $x$, uniformly on $x,\theta\in\R,\,L>0$. We have, for all $x,u,\theta\in\R$,
\beqa\label{eq:Deltavarepproof1}
	\lefteqn{|\Delta\b\varep(x+u,\theta)-
	\Delta\b\varep(x,\theta)|} &&\n
	&\leq&
	\frc1{2\pi}\int\dd x'\dd \theta'\,L\psi_{x\theta}(Lx-Lx',\theta-\theta')\,\Big|
	e^{-\b\varepsilon(x'+u,\theta')}
	-
	e^{-\b\varepsilon(x',\theta')}
	\Big| \\
	&=&
	\frc1{2\pi}\int\dd x'\dd \theta'\,L\psi_{x\theta}(Lx-Lx',\theta-\theta')\,\Big|
	e^{-\beta(x'+u,\theta')}e^{-\Delta\b\varepsilon(x'+u,\theta')}
	-
	e^{-\beta(x',\theta')}e^{-\Delta\b\varepsilon(x',\theta')}
	\Big| \n
	&\leq&
	\frc1{2\pi}\int\dd x'\dd \theta'\,L\psi_{x\theta}(Lx-Lx',\theta-\theta')e^{-\beta(x',\theta')}\,\Big|
	e^{-\Delta\b\varepsilon(x'+u,\theta')}
	-
	e^{-\Delta\b\varepsilon(x',\theta')}
	\Big| + \frc{e^{\beta_{+}}\varphi_{\rm tot} R(u)}{2\pi}.\no
\eeqa
Here we used a.e.~uniform continuity \eqref{eq:uniformcontinuity}, the bound \eqref{eq:bounduniformdeltavarep}, finiteness of the total scattering shift \eqref{eq:integralphi} and continuity of $\psi_{x\theta}$ (thus avoiding the possibly non-empty measure-zero set of discontinuities) to evaluate:
\beqa
	\lefteqn{
	\frc1{2\pi}\int\dd x'\dd \theta'\,L\psi_{x\theta}(Lx-Lx',\theta-\theta')e^{-\Delta\b\varepsilon(x'+u,\theta')}\,\Big|
	e^{-\beta(x'+u,\theta')} - e^{-\beta(x',\theta')}
	\Big| 
	} &&\n
	&\leq &
	\frc{R(u)}{2\pi}\int\dd x'\dd \theta'\,L\psi_{x\theta}(Lx-Lx',\theta-\theta')e^{-\Delta\b\varepsilon(x'+u,\theta')} \n
	&\leq&
	\frc{e^{\beta_{+}}R(u)}{2\pi}\int\dd x'\dd \theta'\,L\psi_{x\theta}(Lx-Lx',\theta-\theta')= \frc{e^{\beta_{+}}\varphi_{\rm tot} R(u)}{2\pi}.
\eeqa
Recall that $\lim_{u\to0} R(u) = 0$. Continuing, we make use of the fact that $1-e^{-u}<u$ for every $u>0$. From the bounds \eqref{eq:boundbetasup},  \eqref{eq:bounduniformdeltavarep} and \eqref{eq:integralphi}, as well as the definition \eqref{eq:defDproof}, we have
\beqa
	\lefteqn{|\Delta\b\varep(x+u,\theta)-
	\Delta\b\varep(x,\theta)|} &&\n
	&\leq&
	\frc1{2\pi}\int\dd x'\dd \theta'\,L\psi_{x\theta}(Lx-Lx',\theta-\theta')e^{-\beta(x',\theta')}\max\big(e^{-\Delta\b\varepsilon(x'+u,\theta')},e^{-\Delta\b\varepsilon(x',\theta')}\big)\,\times\n && \,\Big|
	1
	-
	e^{-|\Delta\b\varep(x'+u,\theta')-
	\Delta\b\varep(x',\theta')|}
	\Big| + \frc{e^{\beta_{+}}\varphi_{\rm tot} R(u)}{2\pi}\n
	&\leq&
	D \sup_{(x,\theta)\in\R^2} |\Delta\b\varep(x+u,\theta)-
	\Delta\b\varep(x,\theta)|
	+ \frc{e^{\beta_{+}}\varphi_{\rm tot} R(u)}{2\pi}.
	\label{eq:deltavarepbla}
\eeqa
By the first inequality of \eqref{eq:Deltavarepproof1} and using \eqref{eq:bounduniformvarep}, $\sup_{(x,\theta)\in\R^2} |\Delta\b\varep(x+u,\theta)-\Delta\b\varep(x,\theta)|\leq 2e^{\beta_{+}-\beta_{-}}\varphi_{\rm tot}/(2\pi)$ is finite. Since $D<1$ (Eq.~\eqref{eq:defDproof}), we find:
\beq
	\sup_{(x,\theta)\in\R^2}|\Delta\b\varep(x+u,\theta)-
	\Delta\b\varep(x,\theta)| \leq
	\frc1{1-D}\frc{e^{\beta_{+}}\varphi_{\rm tot} R(u)}{2\pi}
\eeq
(note that the right hand-side does not depend on $L$). Therefore $\Delta\b\varep(x,\theta)$ is continuous in $x$, uniformly for $x,\theta\in\R$ and $L>0$.

By boundedness \eqref{eq:bounduniformdeltavarep}, $e^{-\Delta\b\varep(x,\theta)}$ is also continuous in $x$, uniformly for $x,\theta\in\R$ and $L>0$: this is obtained by evaluating $|e^{-\Delta\b\varep(x+u,\theta)}-e^{-\Delta\b\varep(x,\theta)}|\leq e^{\beta_{+}}(1-e^{-|\Delta\b\varep(x+u,\theta)-\Delta\b\varep(x,\theta)|})\leq e^{\beta_{+}}|\Delta\b\varep(x+u,\theta)-\Delta\b\varep(x,\theta)|$. Thanks to a.e.~uniform continuity \eqref{eq:uniformcontinuity} and boundedness \eqref{eq:boundbetasup}, then $e^{-\Delta\b\varep(x,\theta)} e^{-\beta(x,\theta)} = e^{-\b\varep(x,\theta)}$ is continuous in $x$ uniformly for a.e.~$(x,\theta)\in\R^2$ and all $L>0$.

Now define
\beq
	\varep^{\infty}(x,\theta) =
	\beta(x,\theta)-\frac{1}{2\pi}\int\dd x'\dd \theta'\,L\psi_{x\theta}(Lx-Lx',\theta-\theta')e^{-\b\varepsilon(x,\theta')}
	=
	\beta(x,\theta)-\int\frac{\dd \theta'}{2\pi}\,\varphi(\theta-\theta')e^{-\b\varepsilon(x,\theta')}.
\eeq
Using \eqref{eq:boundbetasup}, \eqref{eq:integralphi} and \eqref{eq:bounduniformvarep} along with \eqref{eq:defDproof} we find that $e^{-\varep^\infty(x,\theta)}$ is uniformly bounded,
\beq\label{eq:bounduniformvarepinfinity}
	\sup_{(x,\theta)\in\R^2,L>0} e^{-\varep^\infty(x,\theta)} \leq e^{-\beta_{-}+D}.
\eeq
Then
\beqa
	|\b\varep(x,\theta)-\varep^\infty(x,\theta)|
	&\leq&
	\frc1{2\pi}
	\int\dd x'
	\dd\theta' \,
	L\psi_{x\theta}(Lx-Lx',\theta-\theta')\,|
	e^{-\b\ep(x',\theta')}- e^{-\b\ep(x,\theta')}|\n
	&\leq &
	\frc{1}{2\pi}
	\int\dd x'\dd\theta' \,
	L\psi_{x\theta}(Lx-Lx',\theta-\theta')\,
	\sup_{u\in [x-\ep/L,x+\ep/L]}|
	e^{-\b\ep(u,\theta')}- e^{-\b\ep(x,\theta')}|
	\n
	&= &
	\frc{1}{2\pi}
	\int \dd\theta' \,
	\varphi(\theta-\theta')\,
	\sup_{u\in [x-\ep/L,x+\ep/L]}|
	e^{-\b\ep(u,\theta')}- e^{-\b\ep(x,\theta')}|
	\n
	&\to& 0\quad \mbox{uniformly for a.e.~$x\in\R$ and all $\theta\in\R$} \qquad (L\to\infty)
\eeqa
where the last line holds thanks to $L,x,\theta$-uniform continuity of $e^{-\b\varep(x,\theta)}$ as well as the condition \eqref{eq:integralphi}. Hence,\beq
	\lim_{L\to\infty} \Bigg|\b\varep(x,\theta) -
	\Bigg(
	\beta(x,\theta)-\int\frac{\dd \theta'}{2\pi}\,\varphi(\theta-\theta')e^{-\b\varepsilon(x,\theta')}
	\Bigg)\Bigg|
	= 0\quad \mbox{uniformly for a.e.~$x\in\R$ and all $\theta\in\R$}.
\eeq

The above shows that the equation
\begin{equation}\label{eq:proofTBA}
    \varepsilon(x,\theta)=\beta(x,\theta)-\int\frc{\dd\theta'}{2\pi}\,\varphi(\theta,\theta')e^{-\varepsilon(x,\theta')}
    \qquad \mbox{a.e. }x\in\R,\;\forall\,\theta\in\R
\end{equation}
has at least one solution. It also shows that, as $L\to\infty$, $\b\varep(x,\theta)$ approaches a.e.~a set of solutions (in the sense that the additive correction term in the equation as shown tends to zero $x,\theta$-uniformly). We now show that the solution is in fact unique. Suppose $\varep_1$ and $\varep_2$ are two solutions  (for some fixed value of the $x$ argument), and that both satisfy the bound \eqref{eq:bounduniformvarep}. Then
\beqa
	|\varep_1(x,\theta)-\varep_2(x,\theta)|
	&\leq&
	\int \frc{\dd\theta'}{2\pi} \varphi(\theta,\theta')
	\Big|e^{-\varep_1(x,\theta')}-e^{-\varep_2(x,\theta')}\Big|\n
	&\leq& 
	e^{\beta_+-\beta_{-}}\int \frc{\dd\theta'}{2\pi} \varphi(\theta,\theta')
	\Big|1-e^{-|\varep_1(x,\theta')-\varep_2(x,\theta')|}\Big|\n
	&\leq&
	e^{\beta_+-\beta_{-}}\int \frc{\dd\theta'}{2\pi} \varphi(\theta,\theta')
	|\varep_1(x,\theta')-\varep_2(x,\theta')|\n
	&\leq&
	D\sup_{\theta\in\R} |\varep_1(x,\theta)-\varep_2(x,\theta)|
\eeqa
where $\sup_{\theta\in\R} |\varep_1(x,\theta)-\varep_2(x,\theta)|\leq 2D$ by the bound implied by the first line, thus is finite. As $D<1$ the last line implies $|\varep_1(x,\theta)-\varep_2(x,\theta)|=0$ for all $x,\theta\in\R$.

Therefore, we conclude that the limits $\lim_{L\to\infty}\b\varep(x,\theta)$ and $\lim_{L\to\infty}e^{-\b\varep(x,\theta)}$ exist uniformly for a.e.~$x\in\R$ and all $\theta\in\R$ and correspond to the unique solution that satisfy \eqref{eq:proofTBA}. Thus the limit $\lim_{L\to\infty}f_L$ also exists, thanks to \eqref{eq:fLproof_2} (boundedness of $e^{-\b\varep(x,\theta)}$ guarantees that the set of $x$'s of measure zero where the limit might not exist, does not affect the result). From \eqref{eq:bounduniformvarep} and \eqref{eq:defDproof} the inequality \eqref{eq:boundsolution} holds. This completes the proof of the theorem.
\eproof

\section{Conclusion}\label{sectconclu}
In this paper we introduced a new class of classical integrable models. They are constructed from Galilean non-interacting particles using a canonical transformation determined by an explicit generating function. The resulting dynamics of the particles is a ``tracer dynamics": If it is sufficiently far away from other particles, a particle evolves with its asymptotic momentum $\theta$. If two particles scatter they both accumulate a Wigner time delay, which effectively displaces the trajectories of both particles according to their scattering phase-shift. This construction has a lot of freedom, with a freely chosen function $\psi(x,\theta)$ (under some weak constraints) which describes the trajectories of the particles during scattering. Thus, this allows us to construct integrable models with any phase-shift and any interaction range. Here, the constraint we impose on $\psi(x,\theta)$ mean that the phase shift is non-negative and that the interaction range is finite, but the idea is more general, as described in \cite{DHY23b}.

We showed that the resulting models are Liouville integrable with finite-range interaction and that their scattering is elastic and factorised. We showed that they can be obtained as generalized \ttbar- deformations of the free particle model, thus solving the problem of  their locality properties in this context, at least for deformations giving rise to positive phase shifts. Furthermore, we proved that their generalised Gibbs ensembles (GGEs) are described by the usual thermodynamic Bethe ansatz in infinite volume, and by a similar set of equations even in finite volumes, and this, for any space-varying generalised temperatures; in infinite volume we recover the local density approximation. We believe this new class of integrable models very clearly illustrates the structures of many-body integrability.

As a next natural step one could consider the time evolution from slowly-varying GGEs, which is expected to be given by generalized hydrodynamics (GHD); we provide arguments for this in \cite{DHY23b}. Since GHD is a universal theory, these dynamical systems should reproduce all features of GHD, including diffusive and dispersive corrections and correlations. Proving the emergence of GHD rigorously is a complicated problem and has only been accomplished in very special models (most notably hard rods models \cite{Boldrighini1983,ferrari2022hard}); the relatively simple nature of the present family of models, in particular the explicitly known $x\to y$ map, which allows to trivialise the dynamics, offers the prospect of a rigorous proof of the emergence of GHD using a similar strategy as that used for hard rods, and of an understanding of the hydrodynamic scale for a wide range of two-body scattering shifts. Another question is that of the processes of relaxation from far-from-GGE states, which again should be accessible in our models.

For negative phase shifts, the present construction must be modified: as $\psi_{x\theta}(x,\theta)$ is not a non-negative function, the transformation generated by $\Phi(x,\theta)$, Eqs.~\eqref{yeq}, \eqref{peq}, are not necessarily invertible. One may still make physical sense of this situation in various ways. One of them is by borrowing Feynman's idea, and considering processes where pairs of particles and anti-particles are created and annihilated; this is quite novel in the context of classical particles. This is explained in \cite{DHY23b}. We believe that this interpretation can be made rigorous, giving classical integrable models represented by particle non-conserving Hamiltonians on the classical `Fock space' $\bigotimes_N \mathbb{R}^{2N}$. We hope to come back to this in future works.


Finally, other interesting directions of research would be to try to find similar explicit implementations of \ttbar-deformed models in the quantum case; investigate how the present classical models can be useful for numerical implementations, either to simulate the GHD equation or as a tool for numerical preparation of soliton gases \cite{suret2023soliton} (see the discussion in \cite{DHY23b}); and study the inclusion of external force terms, expected to be of the form, for an energy function $E(x,\theta)$,
\begin{align}
    E_\theta(x_i,\theta_i) &=
    \dot x_i + \sum_{j\neq i}
    \psi_{x\theta}(x_i-x_j,\theta_i-\theta_j)(\dot x_i-\dot x_j)\\
    -E_x(x_i,\theta_i) &=
    \dot \theta_i + \sum_{j\neq i}
    \psi_{x\theta}(x_i-x_j,\theta_i-\theta_j)(\dot \theta_i-\dot \theta_j)
\end{align}
(which conserves the total energy $\sum_i E(x_i,\theta_i)$).

\section{Acknowledgements}
BD is grateful to Herbert Spohn for a discussion about the scattering theory of classical systems and pointing out the paper \cite{hubacherClassical1989}. FH acknowledges funding from the faculty of Natural, Mathematical \& Engineering Sciences at King's College London. BD was supported by the Engineering and Physical Sciences Research Council (EPSRC) under grant EP/W010194/1.
\appendix

\section{Details of the use of the matrix-tree theorem}
\label{app:matrixtree}

This involves directed spanning forests for the vertices labelled by $i\in\{1,\ldots,N\}$. A directed spanning forest is a directed graph that contains all these vertices, and that is a disjoint union of directed trees. A directed tree is a tree where each vertex has at most one incoming edge, and exactly one vertex has no incoming edge -- the latter vertex is the root of the tree.
\begin{theorem}[Matrix tree theorem]
    Let $L_{ij}=\delta_{ij}
	\sum_{k\neq i}^NK_{ik}
	-
	(1-\delta_{ij})K_{ij}$ be a Laplace matrix of size $N$ and $\alpha\subset \{1,\cdots,N\}$.  $K_{ij}\in\mathbb{R}$ is a weight matrix that is not necessarily symmetric. The principal minor $\mathcal{L}(\alpha|\alpha)$ obtained by removing rows and columns indexed by the elements of $\alpha$ then admits an expansion as \cite{doi:10.1137/0603033}:
    \begin{equation}
        \mathcal{L}(\alpha|\alpha)=\sum_{F\in\mathcal{F}_\alpha}\prod_{\langle jk\rangle\in F}K_{jk},
    \end{equation}
    where $\mathcal{F}_\alpha$ is the set of directed spanning forests with root set $\alpha$, and $\langle jk\rangle$ is a directed edge from the vertex $j$ to the vertex $k$ ($\prod_{\langle jk\rangle\in F}K_{jk} = 1$ if $F$ has no edge, and $\mathcal L(\emptyset|\emptyset)=0$).
\end{theorem}
With the matrix tree theorem, the Gaudin determinant can be written as
\begin{equation}
      |\Gamma|=\sum_{\alpha\subset\{1,\cdots,N\}}\sum_{F\in\mathcal{F}_\alpha}\prod_{\langle jk\rangle\in F}\psi_{x\theta}(x_{jk},\theta_{jk}),
\end{equation}
and accordingly the partition function becomes
\begin{equation}\label{eq:gge3}
    Z=\sum_{N=0}^\infty\frac{1}{(2\pi)^NN!}\int_{\mathcal V}\prod_{j=1}^N\dd x_j\dd \theta_j\,e^{-\sum_{i=1}^N \beta(x_i/L,\theta_i)}\sum_{\alpha\subset\{1,\cdots,N\}}\sum_{F\in\mathcal{F}_\alpha}\prod_{\langle jk\rangle\in F}\psi_{x\theta}(x_{jk},\theta_{jk}).
\end{equation}

The summation can be carried out as follows. For fixed $N$, the summation generates all the forests that consist of trees, each of which contains one vertex in $\alpha\subset\{1,
\cdots,N\}$ (so that the total number of trees in a given forest is $|\alpha|$). Each tree carries a numerical value that can be computed by assigning $e^{-\beta(x_i/L,\theta_i)}$ to each vertex labeled by $i$, and $\psi_{x\theta}(x_{jk},\theta_{jk})$ to the edge that connects vertices $j$ and $k$. Note that $\psi_{x\theta}(x_{jk},\theta_{jk})$ is symmetric with respect to the two indices, hence we can omit the directions and consider non-directed trees. This gives rise to the following diagrammatic expression of the partition function
\begin{figure}[ht!]
\begin{align}
    Z&=1+\sum_{(\theta_1,x_1)}\quad\raisebox{-12pt}{\includegraphics[width=.015\linewidth]{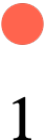}}\n
    &\quad+\frac{1}{2!}\sum_{(\theta_1,x_1)}\sum_{(\theta_2,x_2)} \left(  \raisebox{-15pt}{\includegraphics[width=.25\linewidth]{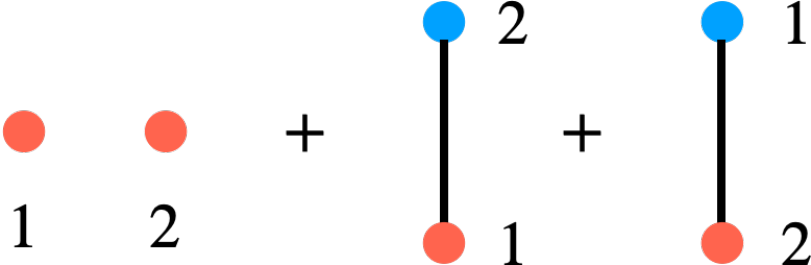}}
   \right)\n   
   &\quad+\frac{1}{3!}\sum_{(\theta_1,x_1)}\sum_{(\theta_2,x_2)}\sum_{(\theta_3,x_3)} \left(  \raisebox{-90pt}{\includegraphics[width=.6\linewidth]{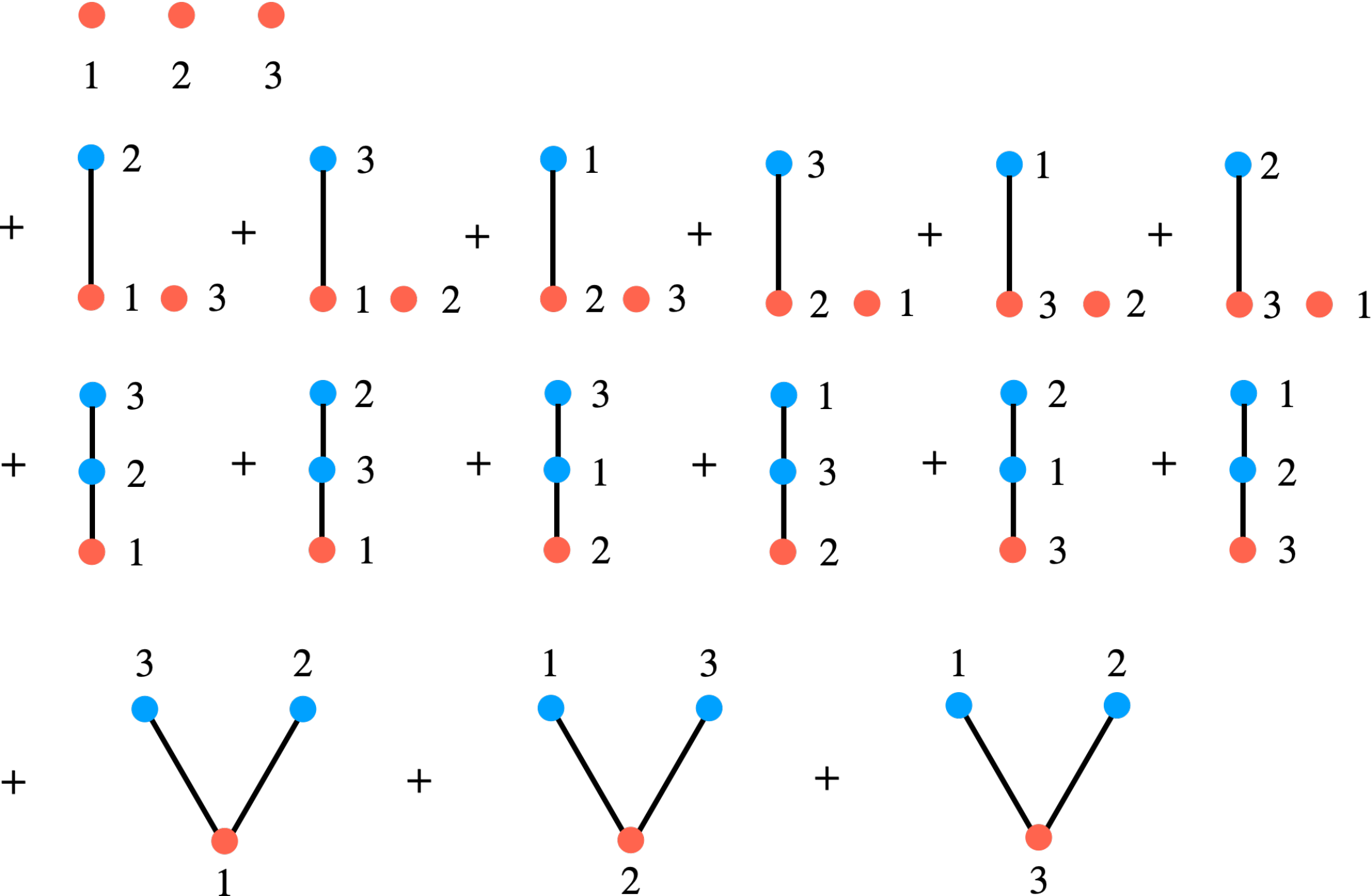}}
   \right)\n
   &\quad+\cdots,
   \end{align}
\end{figure} 
\begin{flushleft}
where $\sum_{(\theta,x)}=\int_{[-rL,rL]\times\mathbb{R}}\dd x\dd \theta/(2\pi)$ and the ellipsis contains terms with higher numbers of integrals. The integration on the phase space then yields
\newpage
\end{flushleft}

\begin{figure}[ht!]
\vspace{-0.5cm}
\begin{equation}
    Z=\,\raisebox{-24.2pt}{\includegraphics[width=0.7\linewidth]{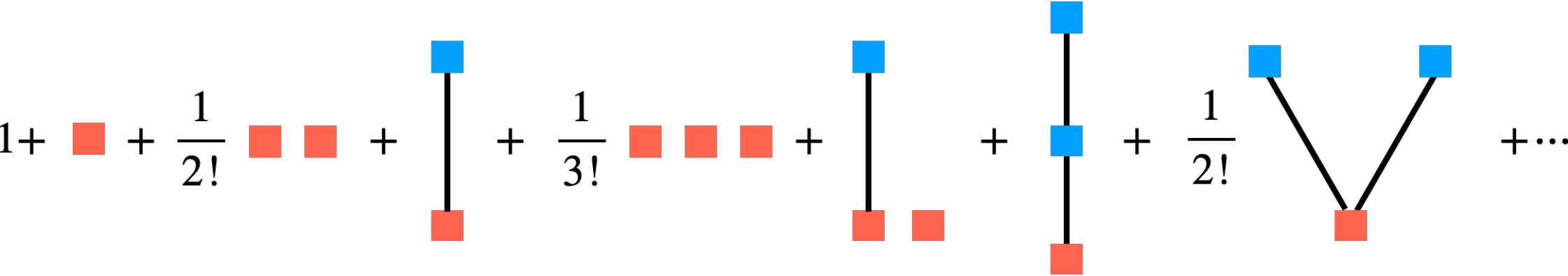}},\label{Ztrees}
\end{equation}
   \end{figure}
   \vspace{-0.5cm}
\begin{flushleft}
where a blue box refers to a vertex integrated over the phase space. We have now lost both directions and vertex labels, and so we consider ``types" of trees (unoriented, unlabelled trees with $N$ vertices). In general, a forest made of $M$ different types of trees, each type $m$ appearing with multiplicity $n_m$, with $\sum_m n_m = N$ vertices, has value
\end{flushleft}
\begin{equation}
    \frac{1}{n_1!}V_1^{n_1} \frac{1}{n_2!}V_2^{n_2}\cdots\frac{1}{n_M!}V_M^{n_M},
\end{equation}
where $V_m$ is the value of the tree of type $m$. Here $1/n_m!$ is simply a symmetry factor for the copies of the tree of type $m$. A tree of type $m$ may have further  internal symmetry factors. A tree with $N$ vertices has $N-1$ edges, and the exact expression for a tree of type $m$ with edges $e_j$ (sets of two vertices) has expression
\begin{equation}
    V_m=\frac{1}{S_m}\frc1{(2\pi)^N}\int_{[-rL,rL]^N\times\mathbb{R}^N}\prod_{j=1}^N\dd x_j\dd \theta_j\,e^{-\sum_{i=1}^N\beta(x_i/L,\theta_i)}\prod_{j=1}^{N-1}\psi_{x\theta}(x_{{\rm e}_j},\theta_{{\rm e}_j}).
\end{equation}

such as in the last term of \eqref{Ztrees}, which is explicitly
\begin{equation}
    V_m=\frac{1}{2!}\frc1{(2\pi)^3}\int_{[-rL,rL]^3\times\mathbb{R}^3}\prod_{j=1}^3\dd x_j\dd \theta_j\,e^{-\sum_{i=1}^3\beta(x_i/L,\theta_i)}\psi_{x\theta}(x_{12},\theta_{12})\psi_{x\theta}(x_{13},\theta_{13}).
\end{equation}
Taking a further summation over $M$, the partition function turns into an exponential of the sum over all the {\it connected trees with an internal symmetry factor}:
\begin{align}\label{eq:gge3_2}
    Z&=\sum_{M=0}^\infty \sum_{\substack{n_1,n_2,\cdots\\ \sum_mn_m=M}}\frac{1}{n_1!}V_1^{n_1} \frac{1}{n_2!}V_2^{n_2}\cdots\frac{1}{n_M!}V_M^{n_M}=\exp\left(\sum_{j=1}^\infty V_j\right) \n
    &=\exp\left[\int_{[-rL,rL]\times\mathbb{R}}\frac{\dd x\dd \theta}{2\pi}Y(x,\theta)\right],
\end{align}
where $Y(x,\theta)$ is the generating function of trees (i.e. sum of all the values of the trees) that are rooted at $(x,\theta)$ \cite{10.1007/978-981-13-2179-5_6}. 

\section{Thermodynamic Bethe ansatz from a saddle-point analysis}

In this section we go beyond the GGE states discussed in Section \ref{ssectTBA} and study local equilibrium states using large deviation theory. Specifically we study measures of the form:
\begin{align}
    \dd{\varrho} &= \frac{1}{Z} \bigoplus_{N=0}^\infty \dd^N{\bs x}\dd^N{\bs p} e^{-\int\dd{x}\dd{\theta}\beta(x/L,\theta) q_\theta(\bs x, \bs p; x)}
\end{align}
in the large-scale limit $L \to \infty$. Note that $\dd^N{\bs x}\dd^N{\bs p} = \dd^N{\bs y}\dd^N{\bs \theta}$ since they are related by a canonical transformation. Using the expressions for the charge densities \eqref{qatheta} we can write this as:
\begin{align}
    \dd{\varrho} &= \frac{1}{Z} \bigoplus_{N}\dd^N{\bs y}\dd^N{\bs \theta} e^{-\sum_{i=1}^N \beta(x_i(\bs y, \bs \theta)/L,\theta_i)}
\end{align}
This is almost the distribution of non-interacting particles, however the positions of particles are shifted due to the presence of other particles. Let rescale $x=L\bar{x}$ and $y=L\bar{y}$ coordinates by $L$ (i.e. going to macroscopic coordinates) and write \eqref{cba1} in the following way:
\begin{align}
    \bar{y}_i(\bs \bar{x}, \bs \theta) &= \bar{x}_i + \frac{1}{L} \sum_{j\neq i} \psi_\theta(\bar{x}_i-\bar{x}_j,\theta_i-\theta_j) = \bar{x}_i + \int\dd{\bar{x}'}\dd{\theta'} \rho_{\rm e}(\bar{x}',\theta') \psi_\theta(L(\bar{x}_i-\bar{x}'),\theta_i-\theta')\\
    &= Y[\rho_{\rm e}](\bar{x}_i,\theta_i)
\end{align}
By this we see that $\bar{y}_i$ is determined only by $\bar{x}_i$ and a mean field coupling to the empirical quasi-particle density:
\begin{align}
    \rho_{\rm e}(\bar{x},\theta) &= \frac{1}{L} \sum_i \delta(\bar{x}-\bar{x}_i)\delta(\theta-\theta_i)
\end{align}

Note that $\bar{x} \to Y[\rho_{\rm e}](\bar{x},\theta)$ is invertible for fixed $\rho_{\rm e}$ and $\theta$ since:
\begin{align}
    \frac{\dd}{\dd\bar{x}} Y[\rho_{\rm e}](\bar{x},\theta) = 1 + \int\dd{\bar{x}'}\dd{\theta'} \rho_{\rm e}(\bar{x}',\theta') \psi_{x\theta}(L(\bar{x}-\bar{x}'),\theta-\theta') > 0.
\end{align}
Denote its inverse for fixed $\rho_{\rm e}$ and $\theta$ by $X[\rho_{\rm e}](\bar{y},\theta)$ and write:
\begin{align}
    \dd{\varrho} &= \frac{1}{Z} \bigoplus_{N}\dd^N{\bar{\bs y}}\dd^N{\bs \theta} \prod_{i=1}^N e^{-\beta(X[\rho_{\rm e}](\bar{y}_i,\theta_i),\theta_i)}\\
    &= \frac{1}{Z} \bigoplus_{N}\dd^N{\bar{\bs y}}\dd^N{\bs \theta} \exp\left(-L \int\dd{\bar{y}}\dd{\theta} \rho_{\rm e,y}(\bar{y},\theta) \beta(X[\rho_{\rm e}](\bar{y},\theta),\theta)\right)\\
    &= \frac{1}{Z} \mathcal{D}[\rho_{\rm e,y}] \exp\left(LS[\rho_{\rm e,y}] - L \int\dd{\bar{y}}\dd{\theta} \rho_{\rm e,y}(\bar{y},\theta) \beta(X[\rho_{\rm e}](\bar{y},\theta),\theta)\right)
\end{align}
where we defined $\rho_{\rm e,y}(\bar{y},\theta) = \frac{1}{L} \sum_i \delta(\bar{y}-\bar{y}_i)\delta(\theta-\theta_i)$ and in the last step changed the measure from $\dd^N{\bar{\bs y}}\dd^N{\bs \theta}$ to $\mathcal{D}[\rho_{\rm e,y}] e^{LS[\rho_{\rm e,y}]}$, which introduces the entropy $S[\rho_{\rm e,y}] = -\int\dd{\bar{y}}\dd{\theta} \rho_{\rm e,y}(\bar{y},\theta) \log \rho_{\rm e,y}(\bar{y},\theta)$ (i.e. the number of microstates per macrostate) in the change of measure. Note that $\rho_{\rm e}(\bar{x},\theta)$ and $\rho_{\rm e,y}(\bar{y},\theta)$ are related via a push-forward $\rho_{\rm e,y} = Y[\rho_{\rm e}]_* \rho_{\rm e}$, i.e.:
\begin{align}
    \rho_{\rm e,y}(\bar{y},\theta) &= \rho_{\rm e}(X[\rho_{\rm e}](\bar{y},\theta),\theta)\frac{\dd{X}[\rho_{\rm e}](\bar{y},\theta)}{\dd{\bar{y}}}
\end{align}
By the large deviation principle we expect that as $L \to \infty$ the measure $\dd{\varrho}$ will become concentrated on the minimizer of the rate function:
\begin{align}
    I[\rho_{\rm e,y}] &= \int\dd{\bar{y}}\dd{\theta} \rho_{\rm e,y}(\bar{y},\theta) \log \rho_{\rm e,y}(\bar{y},\theta) + \int\dd{\bar{y}}\dd{\theta} \rho_{\rm e,y}(\bar{y},\theta) \beta(X[\rho_{\rm e}](\bar{y},\theta),\theta)\\
    &= \int\dd{\bar{x}}\dd{\theta} \rho_{\rm e}(\bar{x},\theta) \log\left(\frac{\rho_{\rm e}(\bar{x},\theta)}{\frac{\dd{Y}[\rho_{\rm e}](\bar{x},\theta)}{\dd{\bar{x}}}}\right) + \int\dd{\bar{x}}\dd{\theta} \rho_{\rm e}(\bar{x},\theta)\beta(\bar{x},\theta)\\
    &= \int\dd{\bar{x}}\dd{\theta} \rho_{\rm e}(\bar{x},\theta) \log\left(\frac{\rho_{\rm e}(\bar{x},\theta)e^{\beta(\bar{x},\theta)}}{1 + \int\dd{\bar{x}'}\dd{\theta'} \rho_{\rm e}(\bar{x}',\theta') \psi_{x\theta}(L(\bar{x}-\bar{x}'),\theta-\theta')}\right)\\
    &\to\int\dd{\bar{x}}\dd{\theta} \rho_{\rm e}(\bar{x},\theta) \log\left(\frac{\rho_{\rm e}(\bar{x},\theta)e^{\beta(\bar{x},\theta)}}{1 + \int\dd{\theta'} \rho_{\rm e}(\bar{x},\theta') \varphi(\theta-\theta')}\right)
\end{align}
Here we used that $\psi_{x\theta}(Lx,\theta)$ approaches $\delta(x)\varphi(\theta)$ as $L\to\infty$. The minimizer of $I[\rho_{\rm e,y}]$ satisfies:
\begin{align}
    0 &= \log\left(\frac{\rho_{\rm e}(\bar{x},\theta)e^{\beta(\bar{x},\theta)}}{1 + \int\dd{\theta'} \rho_{\rm e}(\bar{x},\theta') \varphi(\theta-\theta')}\right) - \int\dd{\theta'} \varphi(\theta-\theta') \frac{\rho_{\rm e}(\bar{x},\theta')}{1 + \int\dd{\theta''} \rho_{\rm e}(\bar{x},\theta'') \varphi(\theta'-\theta'')}
\end{align}
In analogy with the terminology of GHD let us define the `occupation function' $\rho_{\rm n}(\bar{x},\theta) = \frac{2\pi \rho_{\rm e}(\bar{x},\theta)}{1 + \int\dd{\theta'} \rho_{\rm e}(\bar{x},\theta') \varphi(\theta-\theta')}$, which satisfies:
\begin{align}
    \rho_{\rm n}(\bar{x},\theta) e^{- \int\frac{\dd{\theta'}}{2\pi} \varphi(\theta-\theta') \rho_{\rm n}(\bar{x},\theta')} &= 2\pi e^{-\beta(\bar{x},\theta)}\label{equ:LDA_final}
\end{align}
This set of equations decouple for different $x$, which shows the local density approximation.


\printbibliography

\end{document}